\documentclass[12pt]{article}
\usepackage{graphicx}
\usepackage{subfigure}
\usepackage[dvips]{color}

\input epsf.tex

\newcommand{\beq}{\begin{equation}}
\newcommand{\eeq}{\end{equation}}
\newcommand{\be}{\begin{equation}}
\newcommand{\ee}{\end{equation}}
\newcommand{\bea}{\begin{eqnarray}}
\newcommand{\eea}{\end{eqnarray}}

\makeatletter
\renewcommand{\theequation}{\thesection.\arabic{equation}}
\@addtoreset{equation}{section}
\makeatother
\def\href#1#2{#2}
\begin{document}

\baselineskip=15.5pt
\pagestyle{plain}
\setcounter{page}{1}

\begin{titlepage}
\begin{flushleft}
      % \hfill                      {\tt hep-th/1102.****}\\
       \hfill                       FIT HE - 14-02 \\
       \hfill                       %KYUSHU-HET 132 \\
\end{flushleft}
%\vspace*{3mm}

\begin{center}
  {\huge Glueball instability and thermalization   \\ 
   \vspace*{2mm}
driven by dark radiation \vspace*{2mm}
%as Vertex of the Baryon and Other Hadrons
}
\end{center}
%\vspace{5mm}

\begin{center}

\vspace*{2mm}
%\vspace*{5mm}
{\large Kazuo Ghoroku${}^{\dagger}$\footnote[1]{\tt gouroku@dontaku.fit.ac.jp},
${}^{}$Masafumi Ishihara${}^{\ddagger}$\footnote[2]{\tt masafumi@wpi-aimr.tohoku.ac.jp},
Akihiro Nakamura${}^{\S}$\footnote[3]{\tt k3880508@kadai.jp},
%Kouki Kubo${}^{\ddagger}$\footnote[2]{\tt kkubo@higgs.phys.kyushu-u.ac.jp},
%Tomoki Taminato${}^{\ddagger}$\footnote[3]{\tt taminato@higgs.phys.kyushu-u.ac.jp},
%Ingo Kirsch $^c$ \footnote[3]{\tt kirsch@phys.ethz.ch}
%\\

Fumihiko Toyoda${}^{\P}$\footnote[4]{\tt ftoyoda@fuk.kindai.ac.jp}
%Nobuhiro Uekusa\footnote[3]{\tt uekusa@higgs.phys.kyushu-u.ac.jp}
%and Masanobu Yahiro
%\footnote[3]{\tt yahiro2scp@mbox.nc.kyushu-u.ac.jp}
%
}\\

%\vspace*{5mm}
\vspace*{2mm}
{${}^{\dagger}$Fukuoka Institute of Technology, Wajiro, 
Higashi-ku} \\
{%\large 
Fukuoka 811-0295, Japan\\}
%\vspace*{2mm}
{%\large 
${}^{\ddagger}$WPI-Advanced Institute for Materials Research (WPI-AIMR),}\\
{%\large 
Tohoku University, Sendai 980-8577, Japan\\}
%\vspace*{2mm}
%{\em ${}^c$ Institut f\"ur Theoretische Physik, ETH Z\"urich, \\ 
%CH-8093 Z\"urich, Switzerland}
%\vspace*{2mm}
%{%\large 
%${}^{\S}$Department of Physics, Kagoshima University, Korimoto 1-21-35, \\Kagoshima 890-0065, Japan\\}
%\vspace*{2mm}
{%\large 
${}^{\P}$Faculty of Humanity-Oriented Science and
Engineering, Kinki University,\\ Iizuka 820-8555, Japan}

\vspace*{3mm}
\end{center}

\begin{center}
{\large Abstract}
\end{center}
We study glueballs in the holographic gauge theories living in a curved space-time.
The dual bulk is obtained as a solution of the type IIB superstring theory with two
parameters, which correspond to four dimensional (4D) cosmological constant $\lambda$ 
and the dark radiation $C$ respectively.
The theory is in the confining phase for $\lambda <0$ and small $C$, 
then we observe stable glueball states in this
theory. However, the stability of the glueball states
is lost when the density of the dark radiation ($C$) increases and
exceeds a critical point. Above this point, the dark radiation works as the heat bath of the 
Yang-Mills theory since the event horizon appears.
Thus the system is thermalized, and the theory is in 
a finite temperature deconfinement phase, namely in the QGP phase. 
We observe this transition process through the glueball spectra 
which varies dramatically with $C$. We also examined the entanglement entropy of the system
to find a clue of this phase transition and the role of the dark radiation $C$ in the 
entanglement entropy.

\noindent

\vfill
\begin{flushleft}

\end{flushleft}
\end{titlepage}
\newpage

\vspace{1cm}
%%%%%%%%%%%%%%%%%%%%%%%%%%%%%%%%%%%%%%%%%%%%%%%
\section{Introduction}
The 
holographic approach is a
powerful method to study the non-perturbative properties of 
the strong coupling gauge theories \cite{ads1,ads2,ads3}. In this context,
various attempts have been performed to study the properties of the supersymmetric 
Yang Mills (SYM) theory. 
While most of these approaches have been performed for the four dimensional (4D) theory living
in the Minkowski space-time, the analysis has been extended to the theory living in the 
Friedmann-Robertson-Walker (FRW) type space-time \cite{H,GIN1,GIN2,EGR,EGR2,GN13,GNI13}. 
% to find the rich phase structure of the SYM theory.
In this case, two free parameters, the 4D cosmological constant ($\lambda$) and
the dark radiation ($C$), have been introduced in the asymptotic AdS$_{5}$ solution.

Due to the parameter $\lambda$, 
the boundary geometry is changed from the Minkowski to $dS_4 (AdS_4)$
space-time for $\lambda>0~ (\lambda<0)$. Then this solution opens the way to
the holographic approach to the SYM theory in the curved space-time.
In the present case, it has been cleared that 
the dynamical properties of the SYM theory is largely influenced by the 
geometry of the boundary as shown in \cite{H,GIN1} and \cite{GIN2}
for dS$_4$ and AdS$_4$ respectively. 
Especially, in the case of $\lambda<0$ or AdS$_4$ boundary, it has been 
found that the theory is 
in the confining phase \cite{GIN2}. 

As for the dark radiation, on the other hand, it has been introduced in the 
context of the Randall-Sundrum brane-world cosmology \cite{BDEL,Lang}. 
In the context of the brane world model, this term has been regarded as the projection
of the 5D Weyl term on the 4D brane \cite{SMS,SSM}.
From the holographic viewpoint, however, 
this term should be identified with the thermal excitation of SYM fields as observed
in the limit of $\lambda=0$ \cite{EGR,EGR2,GN13,GNI13}.
For $\lambda=0$,
the bulk configuration with $C$ is expressed by the Schwaltzschild-AdS$_5$ by a
redefinition of the radial coordinate, and then we find the Hawking temperature which is proportional
to $C$ \cite{EGR}. It is well known that this configuration is dual to the finite temperature
SYM theory in deconfinement phase.

The dark radiation $C$ therefore competes with the negative $\lambda$ in the dynamics of 
the SYM theory. Namely, $C$ prevents the realization of the confinement phase
which is supported by the negative $\lambda$. This point 
could be expressed by a critical line in the parameter plane of $\lambda -C$
(see the next section) of the quark confinement de-confinement phase transition as has been
discussed in \cite{EGR,GN13,GNI13}. 

In the deconfinement phase, for large $C$, 
the dark radiation would be identified with the thermal
YM fields or the exited gluons as mentioned above. 
{On the other hand, one may wonder what kind of
object is identified with this dark radiation in the confinement phase for small $C$. 
%Is it a fluid of the glueball? 
In order to resolve this point, 
we here study more about this phase transition through the 
glueball spectra since it may be related to the dark radiation in the confinement phase with finite $C$.}

\vspace{.3cm}
Here we should point out another characteristic point observed for negative $\lambda$ case.
It is the second boundary in the infrared side of the bulk as discussed in \cite{GNI13}. 
In the case of zero and small $C$, in the confinement phase, 
there is no special point like horizon in the bulk between 
the two boundaries, which are both described by AdS$_4$. 
It would be related to the fact
that the two AdS$_4$ are connected for $C=0$ at their
3D boundary, which are described by 3D hyperbolic space $H^3$, \cite{KR,M}. 
Then the field operators living on each AdS$_4$ space-time would 
extend to the other AdS$_4$. This behavior of the field operators would be observed
in the bulk in some way, then we expect to be able to see it 
through the holographic analysis.

For the case of $C=0$, we actually find that
the metric of the bulk is symmetric under an inversion transformation
($z=r_0^2/r$) of the fifth coordinate ($r$)
at a point $r=r_0$. 
We call the 4D slice at this point as "domain wall" since the existing region 
of the strings and the branes introduced
as probes to investigate the dynamics of the dual theory is restricted to the range
$r_0<r<\infty$ or $0<r<r_0$.
\footnote{This position of the coordinate has been also
noticed as a node of the wormhole in a slightly different
context of holography for two boundary theories \cite{M}.}
This point has been discussed in \cite{GNI13}.
Then the bulk is separated to two regions by this domain wall. 
Then we expect to find the equivalence of the theory on the boundary $r=\infty$ and
the one at $r=0$ for $C=0$. 

\vspace{.3cm}
For $C\neq 0$, the position of the domain walls is changed depending on the 
quantities to be studied.
The wall for the static string, which is used to see the string tension responsible to 
the quark-confinement, 
and for the entanglement entropy
are shown as such examples. They are at the same point
for $C=0$, then we can see how they vary with increasing $C$.  
After the transition to the deconfinement phase at large $C$, a horizon appears and
the second boundary is hidden behind the horizon. 

\vspace{.3cm}
{On the other hand, the dual of the glueballs are examined through the 
fluctuation of the bulk fields or as rotating closed string configurations. When we observe 
the fluctuation of the fields, the domain wall seems to be disappearing since the 
wave-function of the fluctuation spreads out all over the bulk.
However, we could see that the center of the wave-function of the glueball state
and the classical configuration of rotating closed string are just on the domain wall.
In this sense, the role of the domain wall is altered in the glueball case. The glueballs
are attracted at the domain wall and they could spread as the quantum fluctuations as shown 
below. The situation depends on the dark radiation $C$. The glueballs can be observed 
in both theories on the opposite side boundaries. 
}
%%***********************************

The outline of this paper is as follows.
In the next section, the bulk solutions for our holographic model are given, then 
some important points are briefly reviewed. In the section 3, the spectra of glueballs
in the case of $C=0$ are shown as an exact form of analytical solution. 
Also, the spectrum for $C\neq 0$ is estimated by WKB approximation
for the lowest mass of the glueball to see its behavior near the transition point to the deconfinement phase. 
In the Sec. 4, the glueballs with higher mass state are examined by solving the equation
of motion for the rotating closed string with folding form. We could show that
the properties of the solutions 
for $C\neq 0$ are consistent with the one given in the previous section.
In the Sec. 5, the entanglement entropy has been examined and we could find the thermal
limit of the dark radiation part in the deconfinement phase for large volume limit of the
considering minimal surface. Other interesting properties are also given and discussed.
The summary and discussions are given in the final section.
%%%%%%%%%%%%%%%%%%%%%%%%%%%%%%%%%%%%%

\section{Gravity dual of dark energy and dark radiation}

The holographic dual to the large $N$ gauge theory embedded in a space-time with dark energy 
and dark radiation is solved by the gravity on the following form of the metric
\beq\label{10dmetric-2}
ds^2_{10}={r^2 \over R^2}\left(-\bar{n}^2dt^2+\bar{A}^2a_0^2(t)\gamma_{ij}(x)dx^idx^j\right)+
\frac{R^2}{r^2} dr^2 +R^2d\Omega_5^2 \ . 
\eeq
where
\beq\label{AdS4-30} 
    \gamma_{ij}(x)=\delta_{ij}\left( 1+k{\bar{r}^2\over 4\bar{r_0}^2} \right)^{-2}\, , \quad 
    \bar{r}^2=\sum_{i=1}^3 (x^i)^2\, ,
\eeq
and $k=\pm 1,$ or $0$. The arbitrary scale parameter  $\bar{r_0}$ of three spase
is set hereafter as $\bar{r_0}=1$.
The solution is obtained from 10D supergravity
of type IIB theory \cite{EGR,EGR2,GN13,GNI13}. 
Brief review is given in the appendix A.

The resultant solution %of $A$ and $n$ by using $\lambda(t)$. They 
is obtained as
\bea
 \bar{A}&=&\left(\left(1-{\lambda\over 4\mu^2}\left({R\over r}\right)^2\right)^2+\tilde{c}_0 \left({R\over r}\right)^{4}\right)^{1/2}\, , \label{sol-10-1} \\
\bar{n}&=&{\left(1-{\lambda\over 4\mu^2}\left({R\over r}\right)^2\right)^2
         %\left(1-{\lambda+{a_0\over \dot{a}_0}\dot{\lambda} \over 4\mu^2}\left({R\over r}\right)^2\right)
-\tilde{c}_0 \left({R\over r}\right)^{4}\over 
       \sqrt{\left(1-{\lambda\over 4\mu^2}\left({R\over r}\right)^2\right)^2+\tilde{c}_0 \left({R\over r}\right)^{4}}}\, , \label{sol-11-1}
\eea
\beq
\tilde{c}_0=C/(4\mu^2a_0^4)\, , \label{sol-12-1}
\eeq
where the dark radiation $C$ is introduced as an integral constant in solving the equation
of motion (\ref{A1}). On the other hand, the "dark energy" 
(or cosmological term) $\lambda(t)$ is
introduced as follows
\beq
  \left({\dot{a}_0\over a_0}\right)^2+{k\over a_0^2}=\lambda\, \label{bc-3-1}
\eeq
While it is possible to extend $\lambda$ to the time dependent form $\lambda(t)$ as in 
\cite{EGR}, we consider here the case of constant $\lambda$ for simplicity.
{In the following, our discussion would be restricted to the case of
negative constant $\lambda$ and we assume very small time derivative of $a_0(t)$. For the sake
of the justification of our assumption for $a_0(t)$, we should say that the solution
$a_0=$constant is allowed for negative constant $\lambda$ when we take $k=-1$. 
%So in this case, the three dimensional space must be hyperbolic.}

\vspace{.3cm}
The physical meaning of $\lambda$ is clear, however 
the dark radiation $C$ is not familiar. So we explain it here. {For $\lambda=0$, 
the meaning of $C$ is clearly understood. In this case,
the above 5D metric is rewritten into the AdS-Schwartzschild form,
then we find the Hawking temperature $T_H$ as follows \cite{EGR}
\beq\label{temperature-0}
   T_H={\sqrt{2}b_0 \over \pi R^2}\, , \quad b_0=\tilde{c}_0^{1/4} R\, .
\eeq
This implies that the dark radiation
$C$ corresponds to the thermal radiation of SYM fields 
in the Minkowski space-time since $\lambda=0$. It 
is also assured from the VEV of energy momentum tensor that 
the dark radiation corresponds to a perfect fluid of gluons (or radiation) with the
temperature $T_H$ \cite{EGR}.

\vspace{.3cm}
\subsection{Confinement deConfinement Phase transition}

\begin{figure}[htbp]%[H]
\vspace{.3cm}
\begin{center}
\includegraphics[width=8cm]{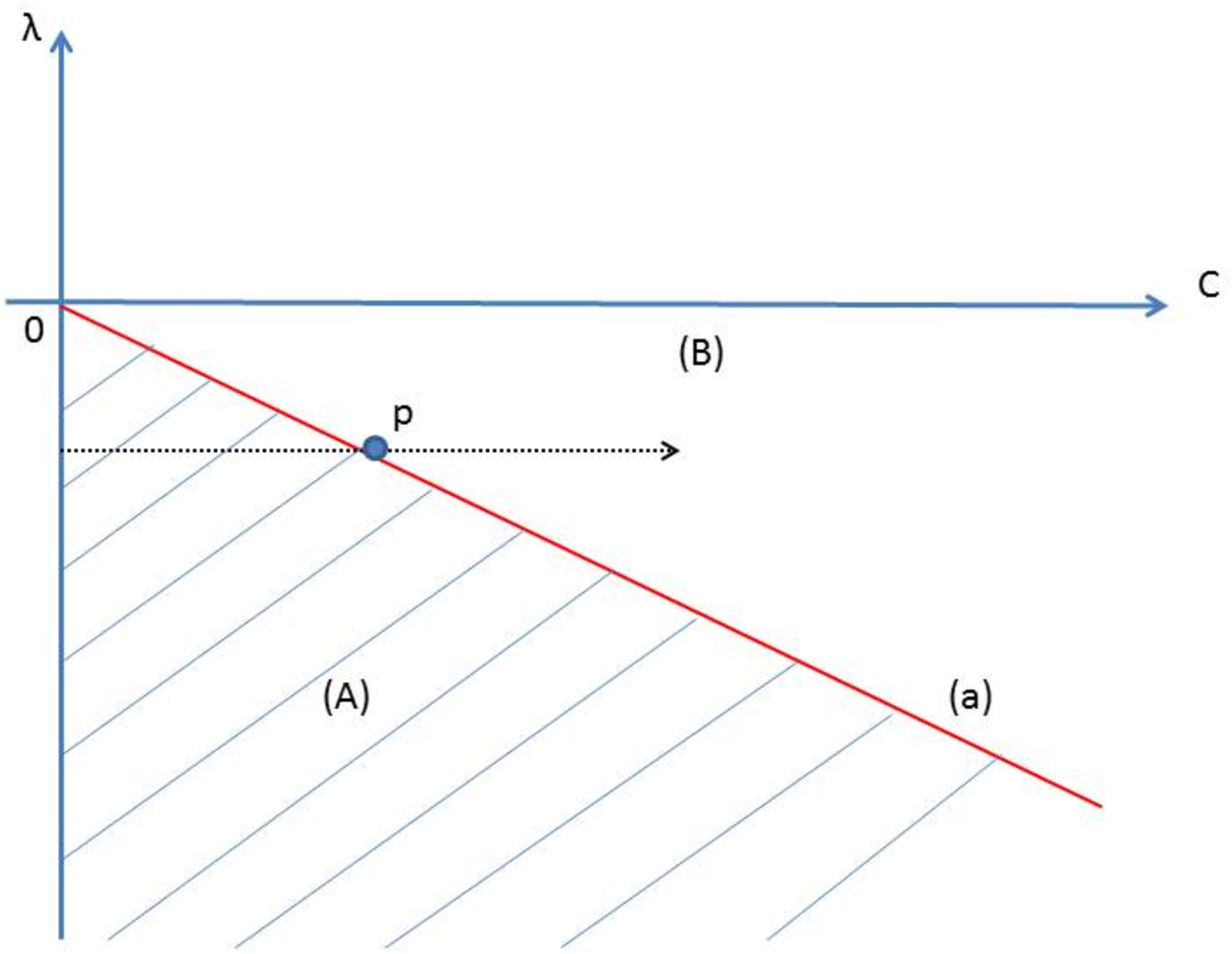}
\includegraphics[width=7cm]{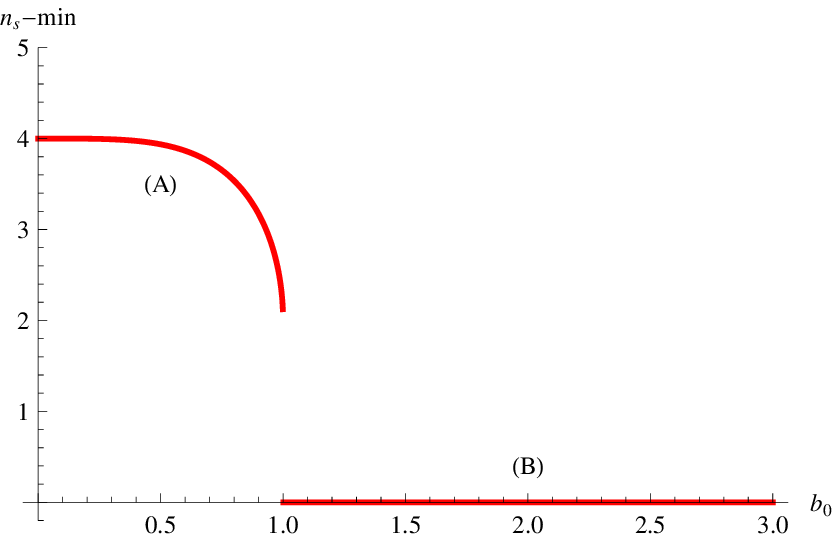}
\caption{\small  Left: Phases (A) (shaded) quark-Confinement and (B) deConfinement
are shown in $\lambda -C$ plane. The critical line $r_0=b_0$ is shown by (a). 
Right: Minimum value of $n_s$ as a function of $b_0$ is shown for $R=r_0=1$ along the vertical line
in the left figure. 
At the point p of the line, we find $b_0=1$ and $n_s(1)=2$.}
\label{Phase-Diagram}
\end{center}
\end{figure}

Therefore, when the dark radiation $C$ is added in some way to the YM theory in the confinement
phase, the confinement force is screened and the phase of the theory is changed to the deconfinement
above a critical value of $C$. This phenomenon has been observed in the $AdS_4$ space-time 
by examining the Wilson loop \cite{EGR}. It has been observed that the SYM system in the $AdS_4$ is in the
confinement phase \cite{GIN1}. 
However. as mentioned above, the phase of the theory is changed to the 
deconfinement one with finite temperature
by adding $C$, which satisfies the condition $b_0>r_0$, where 
\beq
   r_0={R^2\over 2}\sqrt{|\lambda|}\, .
\eeq
In this region, a horizon appears at $r=r_H$ which is given as
\beq\label{temp}
   r_H=\sqrt{b_0^2-r_0^2}\, .
\eeq

\vspace{.2cm}
Then in the $AdS_4$ space-time or for $\lambda<0$
the phase transition occurs at $b_0=r_0$, where the temperature is zero, namely $T_c=0$.
We notice that the temperature is zero in the range of confinement, $0\leq b_0\leq r_0$, and the
temperature appears for $r_0<b_0$, in the deconfinement region. 
The critical point $T_c=0$ is represented therefore by the line $b_0=r_0$ 
in the plane of $\lambda -b_0$ as shown in the Fig. \ref{Phase-Diagram}.

Usualy, this kind of transition is studied through the Hawking-Page transition by using
two independent solutions, confinement solution and the one of the deconfinement. Then 
the Hawking-Page transition has been studied by comparing the free energy of the theories
at a finite temperature for those two bulk solutions \cite{ET,KMMW,arXiv:0908.0407,Sin}. Then we find a critical
temperature as a finite value.

In the present case, the phase transition is examined in terms of the same solution
by varying $b_0$ instead of the temperature.
In this sense, the present model is a new type of holographic model. Our purpose is to
study through this model the phase transition phenomenon in more detail.

\vspace{.2cm}
In this transition, we could consider 
the QCD string tension as the order parameter.
This point is briefly shown below under an assumption
that the time evolution of the universe is very slow, or equivalently for 
$\dot{a}(t)/a_0(t)<<1$. 

\vspace{.3cm}
%\subsection{Confinement and QCD string tension}
\noindent{\bf QCD string tension as an order parameter}

The potential between
quark and anti-quark is studied by the Wilson-Loop for the present case \cite{EGR}. 
It is obtained holographically
from the U-shaped ( in $r-x$ plane) string which is embedded in the bulk
and its two end-points are on 
the boundary. % as studied in \cite{GIN1}.
Supposing a string whose world volume is set in $(t,x)$ plane 
\footnote{Here $x$ denotes one of the three coordinate $x^i$, and we take $x^1$ in the present case.}, 
the energy $E$ 
of this state is obtained as a function of the distance ($L$) between
the quark and anti-quark according to \cite{GIN1}.

Taking the gauge as $X^0=t=\tau$ and $X^1=x^1=\sigma$
for the coordinates $(\tau,~\sigma)$ of string world-volume,
the Nambu-Goto action in the present background (\ref{10dmetric-2}) 
becomes
\beq
   S_{\textrm{\scriptsize NG}}=-{1 \over 2 \pi \alpha'}\int dt d\sigma ~
   {\bar{n}(r)}\sqrt{r'{}^2
        +\left({r\over R}\right)^4 \left({\bar{A}(r)}a_0(t)\gamma(x)\right)^2 } ,
 \label{ng}
\eeq
where
\beq
  \gamma(x)={1\over 1-x^2/4}\, ,
\eeq
and we notice $r'=\partial r/\partial x=\partial r/\partial \sigma$.
Here we notice the metric $\bar{n}$ and $\bar{A}$ 
have time dependence through $\tilde{c}_0$ as given above. 
As mentioned above, we must remember our assumption that
the time derivative of $a_0(t)$ is very small compared to the time scale
of the fields we are considering in the theory. So we could neglect this time-dependence
hereafter.

%%%%%%%%%%%%%%%%%%% FIG %%%%%%%%%%%%%%%%
%%%%%%%%%%%%%%%%%%%%%%%%%%%%%%%%%%%%%%

\vspace{.5cm}
 {From the above action, $S_{\textrm{\scriptsize NG}}$, the string 
configuration for large $x$ is obtained by solving the equation of motion. Using
this solution, we can}
estimate $S_{\textrm{\scriptsize NG}}$. Then the linear potential is obtained when the 
factor $n_s$, which is given as
\beq 
 n_s=\left({r \over R}\right)^2 \bar{A}\bar{n}\, , \label{matter-ns0}\, 
\eeq
has a minimum at some point of $r(=r_D>0)$. Further this minimum must be positive,
$n(r_D)>0$.

In the present case, such a point is found as 
\beq
  r_D=\left( r_0^4-b_0^4\right)^{1/4}\, ,
\eeq
{where $b_o=\tilde{c}_0^{1/4}R$ and
\beq\label{nsmin}
   n_s(r_D)=2{r_0^2\over R^2}\left(1+{r_D^2\over r_0^2}\right)\, .
\eeq}
Then, as shown in the Fig. \ref{Phase-Diagram}, the positive minimum exists in the region, 
\beq
    r_0 \geq b_0\geq 0\, .
\eeq

\begin{figure}[htbp]%[H]
\vspace{.3cm}
\begin{center}
\includegraphics[width=12cm]{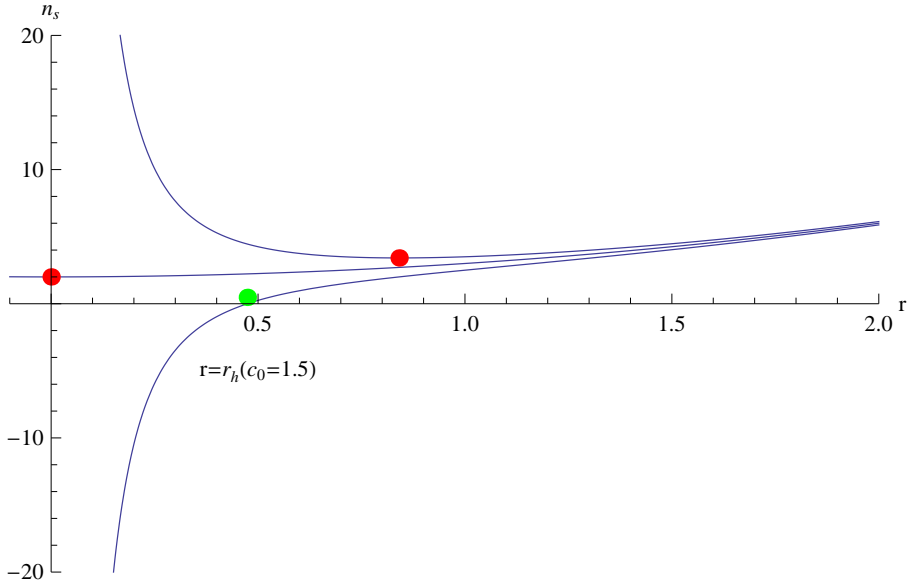}
\caption{{\small  The $n_s$ is plotted as functions of $r$.  
The value of $\tilde{c}_0$ is 0.5, 1, 1.5 from the above.  
There $r_0=1$, $R=1$ and $\lambda=-4 r_0^2/R^4 = -4$.  
The red points are minimum points of $n_s$ for $\tilde c_0=0.5$ and 1.0.  
The green point is the horizon at $\tilde{c}_0=1.5$.    
}}
\label{rns}
\end{center}
\end{figure}

{We notice that there is a gap for the minimum of $n_s$ at
the transition point, $r_D=0$. From (\ref{nsmin}), it is given as
\beq
 n_s(0)=2r_0^2/R^2 \, ,
\eeq 
which is finite. On the other hand, for $b_0>r_0$, the horizon
appears  at $r=r_H$, which is given as
\beq\label{temp}
   r_H=\sqrt{b_0^2-r_0^2}\, .
\eeq 
Then, as shown in the Fig. \ref{rns}, the minimum of $n_s(\geq 0)$  is given at this point as 
\beq
n_s(r_H)=0\, ,
\eeq
for $\tilde{c}_0=(r_0/R)^4>1$ \footnote{Notice that we set $r_0=R=1$ in the Fig. \ref{rns}}. 
A typical case for $\tilde{c}_0=1.5$ is shown in the Fig. \ref{rns}. 
For all range of
$\tilde{c}_0$, the value of minimum of $n_s$ is shown in the Fig. \ref{Phase-Diagram}.
This implies a gap for the string tension at the critical point.} 
This fact implies the first order phase transition.

\vspace{.5cm}
\noindent\underline{ Domain wall} %\cite{GKTT,GNI13}}

We give a comment for the terminology "domain wall" related to the
point $r_D$, where $n_s$ takes its minimum.
We call the 4D slice, which is cut at $r=r_D$ in the bulk,
as a domain wall since the open string configuration introduced
to calculate the string tension as above are prevented to penetrate this wall
\cite{GKTT,GNI13}. Then 
the bulk is separated by this wall to two regions in the case of $0<b_0<r_0$
as shown in the Fig. \ref{domainwall}. We could see that each region is 
dual to the 4D field theory on each boundaries at $r=\infty$ and $r=0$
when we consider the dynamics of the quark and anti-quark. 
However, the situation would depend on the quantity of the corresponding
field theory as shown below. 
Here two kinds of such a wall are shown in the Fig. \ref{domainwall}.  
They depart from the common position with increasing $b_0$ for $C\neq 0$.
Furthermore, the role of the wall is altered for glueball or closed string state.  
This point is explained more in the next subsection.

%noticing the number of the boundaries.
%%%%%%
\subsection{Two Boundaries}\label{sec22}

In performing the analysis in the present model, 
we must notice the point discussed in \cite{GN13}
that the bulk metric given here has two boundaries in the confinement region 
(A) shown in the Fig. \ref{Phase-Diagram},
$\lambda< 0$ and $0\leq b_0<r_0$}.
The two boundaries are found at $r=\infty$ (UV side) and $r=0$ (IR side), 
however there is
no horizon between them. { The reason why the horizon is absent is 
understood as follows.
Each boundary is described by AdS$_4$, and its boundary
is connected to the one of the other AdS$_4$. In this sense, these two boundaries are connected
on their boundaries. It is an interesting problem to see how the two boundary
theories are connected at their boundary. It would be reflected in the bulk,
so we could resolve this point from the holographic approach, however
it is postponed to concentrate on this point and to get a deeper understanding.
Our purpose is to study the role of the dark radiation in the theory on the UV boundary.
As for the IR boundary, we only give understandable few comments.}

%{They might be expressed by the generationg functional of each gauge
%theory, by estimating the partition function of the type IIB supergravity, as follows 
%\bea
%  I_{\rm SUGRA}&=&W_{\rm UV}(\Phi_{0})+W_{\rm IR}(\hat{\Phi}_{0})\, \label{generating} \\
 %   &=&-\log\langle e^{\int d^4x O(x)\Phi_{0}(x)}\rangle_{CFT_1}
  %        -\log\langle e^{\int d^4x \hat{O}(x)\hat{\Phi}_{0}(x)}
   %             \rangle_{CFT_2}\,  , \label{generating-2} 
%\eea
%where $W_{\rm UV}$ ($W_{\rm IR}$) denotes the generating functional of the field theory at UV (IR)boundary with the source field $\Phi_0$ ($\hat{\Phi}_0$), which represents the boundary value of the bulk field $\Phi(x,r)$ at $r=\infty$ ($\hat{\Phi}(x,r)$ at $r=0$).}

\begin{figure}[htbp]%[H]
\vspace{.3cm}
\begin{center}
\includegraphics[width=10cm]{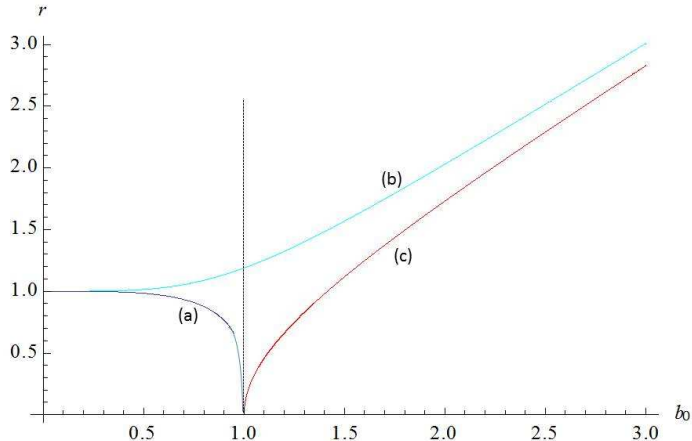}
\caption{{\small  Two dual bulks are shown in in $r-b_0$ plane. 
The domain wall is shown by the curve (a) $r_D=(r_0^4-b_0^4)^{1/4}$
for $r_0=1$. The curve (c) denotes the horizon $r_H=\sqrt{-r_0^2+b_0^2}$}.
The curve (b) $r_{c}=(r_0^4+b_0^4)^{1/4}$ is explained in the section 5.
The vertical line shows the critical line for confinement de-confinement
phase transition.
}
\label{domainwall}
\end{center}
\end{figure}

The holographic situation for these theories on the two boundaries
depends on the parameter $b_0$ as explained below according to the
horizontal axis $b_0$ in the Fig.  \ref{domainwall}.

\vspace{.3cm}
\noindent\underline{\bf $b_0= 0$}

As discussed in \cite{GN13}, 
in the case of $b_0=0$ and $\lambda<0$, the bulk can be separated into two
regions by a border called as "domain wall" which is set at $r=r_0$. 
Then the field theory on each boundary is obtained from each bulk 
separated by this wall. Actually, %we find the same form of metric 
in the present case, we find that the metrics are symmetric around $r=r_0$ under
the transformation $r\to z=R^2/r$. Then we will find the same 
boundary 4D theory. %metric, $g_0=\hat{g}_0$.
In other words, we could observe the same dynamical properties 
of the two field theories at $r=\infty$ and $z=0$. 

\vspace{.3cm}
\noindent\underline{\bf $0<b_0\leq r_0$}

In this region, the two theories
show different properties from each other when the dark radiation $C$ is added.
%\vspace{.5cm}
In fact, in this case, we find different boundary metric, $g_0\neq \hat{g}_0$, 
where $g_0$ ($\hat{g}_0$) denotes the metric on the boundary $r=\infty$ ($r=0$), at the 
two boundaries. 
As a result, the energy momentum tensors are also different in each boundary
as shown below. The point, we should notice, is that the metric
at $r=0$, $\hat{g}_0$, depends on $b_0$. Then
the theories become asymmetric due to the dark radiation. On this point,
we give a brief comment below.

\vspace{.3cm}
\noindent{\bf (i) For $r=\infty$;} The boundary metric $g_0$ 
is not altered by $b_0$
and is given as
\beq\label{UV-metric}
ds^2=-dt^2+a_0^2(t)\gamma_{i,j}dx^idx^j\, .
\eeq
We notice here that the above metric depends only on $\lambda$, but it does not
include the dark radiation 
$C$ or $b_0$. This point is important. The dark
radiation is instead observed as a perfect fluid of the gauge fields as seen
in the energy momentum tensor $\langle  {T}_{\mu\nu}\rangle$ \cite{EGR}.

{The role of this fluid is to screen the confining
force between colored charges. However,
the confining force is overwhelming in the region $0<b_0\leq r_0$, where quark confinement 
is observed.}

\vspace{.3cm}
\noindent{\bf (ii) For $r=0$;} On the other hand, at
$r=0$, the metric $\hat{g}_0$ is deformed by the dark radiation $b_0$. 
It is given as follows \cite{GN13}. 
\beq\label{IR-metric}
ds^2=-\alpha_1dt^2+\alpha_2a_0^2(t)\gamma_{i,j}dx^idx^j\, .
\eeq
\beq\label{IR-metric-2}
   \alpha_1={\left(r_0^4-b_0^4\right)^2 \over \left({r_0}^4+b_0^4\right)r_0^4}\, , \quad
   \alpha_2={{r_0}^4+b_0^4\over r_0^4}\, .
\eeq
The $tt$ component of this metric becomes zero at the critical point, $b_0=r_0$. 
Then the event horizon appears.
%\vspace{.2cm}
As for the energy momentum tensor $\langle {T}_{\mu\nu}^{\rm IR}\rangle$ 
on the boundary $r=0$, which is given in \cite{GN13}, the perfect fluid part disappears.

\vspace{.2cm}
{This fact implies that the dark radiation works as a 4D matter which couples to the gravity
to reform the 4D metric $\hat{g}_{(0)\mu\nu}$. 
Then the bulk would be dual to the pure SYM living in this deformed FRW space-time.
}

\vspace{.3cm}
\noindent\underline{$r_0<b_0$} 

In the region of $r_0<b_0$, 
the IR boundary hides behind the horizon, which appears at $r_H$ for $r_0<b_0$.
Then we consider the region from a horizon $r_H$ to $r=\infty$. 
In this case, the bulk is simply dual only to the theory
at the boundary $r=\infty$. 
{The theory is in the deconfinement phase with finite
temperature. So we expect dynamical properties which are similar to the case of AdS$_5$-Schwartzschild
bulk. However, as shown in the Fig. \ref{domainwall}, the domain wall for the entanglement entropy
appears above the horizon. So we expect a slightly different thermodynamic properties in the
present case compared to the one of AdS$_5$-Schwartzschild finite temperature theory.
}

\vspace{.3cm}
{In the following, we study the phase transition property through glueball spectra and 
entanglement entropy. After the transition to the deconfinement phase at $b_0>r_0$, the theory
we consider is restricted to the one at the boundary $r=\infty$. Then the glueball mass
and entanglement entropy are observed from the UV boundary theory.}

%%%%%%%%%%%%%%%%%%
%\subsection{Rotating string and equations of motion}~

\section{Glueballs from bulk field fluctuations}

In the confinement phase, $0\leq b_0<r_0$, we could expect the existence of
glueball state in the dual theory. It is studied by solving the equation of motion 
of the field fluctuations in the bulk {\cite{ORT}-\cite{BMT},\cite{GTT} }. 
This is performed here by separating to two cases of the parameter $b_0$.

\subsection{$b_0=0$ case }

In this case, $C= 0$, then the corresponding
glueball state is studied 
by solving the field equation of the quantum fluctuation
of the bulk fields in the following background, % (\ref{AdS4-1} )-(\ref{AdS4-2} )

\bea
ds^2_{10}&=&ds^2_{5}+R^2d\Omega_5^2 \nonumber \\ %\label{AdS4-1} \\ 
   ds^2_{5}&=&{r^2 \over R^2}\left(1+{r_0^2\over r^2}\right)^2\tilde{g}_{\mu\nu}dx^{\mu}dx^{\nu}
+\frac{R^2}{r^2} dr^2 \nonumber \\%\label{AdS4-2} 
  \tilde{g}_{\mu\nu}dx^{\mu}dx^{\nu} &=& 
  \left(-dt^2+a_0^2(t)\gamma_{ij}(x)dx^idx^j\right)\, . \label{4dim-metric}
\eea

\vspace{.5cm}
\noindent{\bf  Graviton $2^{++}$; }
As for the glueball spectrum, many attempts have been made by solving the
linearized field equations of bulk field fluctuations in various background configurations.
Here we consider the field equation of the traceless and transverse component
of the metric fluctuation, which is denoted by $h_{ij}$. 
%\textcolor{red}
{%In our both models, its 
Its linearized equation is given 
in the Einstein frame metric as 
\beq\label{graviton-eq}
  {1\over\sqrt{-g}}\partial_M\left(\sqrt{-g}g^{MN}\partial_N h_{ij}\right)=0\,,
\eeq
%\textcolor{red}
where we assumed as $h_{ij}=h_{ij}(x^0,x^i,r)$,} then
$M,N$ are the five dimensional ($(x^0,x^i,r)$) suffices.\footnote{In the string frame metric case, this equation is written as
${1\over\sqrt{-g}}\partial_M\left(\sqrt{-g}
e^{-2\Phi}g^{MN}\partial_N h_{ij}\right)=0$ as given in \cite{Minahan}}
This equation
is equivalent to the massless scalar field equation. As shown in \cite{BMT},
this equation is common to $2^{++},~1^{++}$ and the one of the non-active 
\footnote{Here active means that the dilaton background solution is nontrivial
as in the present case.}
dilaton $0^{++}$ \cite{Gio},
which are dual to the glueball of $F_{\mu\nu}F^{\mu\nu}$.
% and obtained from the metric fluctuations of 11D supergravity with one compactified space dimension. 
While it is usually used to derive the type IIA theory, the NS-NS part
is common with the one of the type IIB theory. 
Then the masses of these three spin states degenerate.

\vspace{.5cm}
By setting as $h_{ij}=p_{ij}\chi(x^{\mu})\phi(r)$, \footnote{
$p_{ij}$ denotes projection operator onto the traceless and transverse components.
}
we impose for the 4D part of the wave-function, $\chi(x^{\mu})$, the following 
eigenvalue equation,
\beq\label{4dads}
   {1\over \sqrt{\tilde{g}_4}}\partial_{\mu}\sqrt{\tilde{g}_4}\tilde{g}^{\mu\nu}
   \partial_{\nu}\chi(x^{\mu})=m^2\chi(x^{\mu})
\eeq
where ${\tilde{g}_4}=-{\rm det}\tilde{g}_{\mu\nu}$.
Then, we get for $\phi(r)$ the following equation,
\bea
   &&\partial_r^2\phi+g_2(r)\partial_r\phi+({R\over r})^4{m^2\over A^2 }\phi=0\, ,\label{glueball-eq-1}\\
  && g_2(r)=\partial_r\left(\log\left[(r/R)^5A^4\right]\right)\, , \label{glueball-eq-2} \\
  && A(r)=1+({r_0\over r})^2\, . \label{glueball-eq-3}
\eea

\vspace{.3cm}
%\noindent{\bf $m^2$ as Eigenvalues of Eq. (\ref{4dads}); for free fields}

{When the equation (\ref{4dads}) is regarded as the one for the free field
in AdS$_4$ space-time, the eigenvalue of $m^2$ has been obtained as 
\beq\label{AdS4-mass}
   m^2=\lambda n(n+3)\, , %\quad n=0,1,2, \cdots
\eeq
for scalar ($n\geq 0$) in \cite{AIS} and for spin two tensor ($n\geq 1$) in \cite{Fr}.
In the following, we could obtain the mass spectra of 
(\ref{AdS4-mass}) for the glueball of $2^{++}$ in AdS$_4$. This fact implies the correctness of the 
holographic approach also to the theory living in AdS$_4$. }

\vspace{.3cm}
\noindent\underline{General solution}

In the present case, we can solve analytically the equation (\ref{glueball-eq-1}).
Changing $r$ to a dimensionless variable $x=r/r_0$, the above equations are rewritten as
\bea
   &&\partial_x^2\phi+g_2(x)\partial_x\phi+{\bar{m}^2\over x^4A^2(x) }\phi=0\, ,\label{glueball-eq-1-2}\\
  && g_2(r)={1\over x}\left(5-{8\over x^2A(x)}\right)\, , \label{glueball-eq-2-2} \\
  && A(x)=1+({1\over x^2})\, , \label{glueball-eq-3-2}
\eea
where $\bar{m}=R^2m/r_0$. This equation (\ref{glueball-eq-1-2}) has two regular singularity
at $x=0$ and $x=\infty$. We solve this equation as follows; Firstly by setting the following form
for $\phi$,
\beq
  \phi=A^ax^bP(x)
\eeq
then it is possible to write the equation for $P(x)$ as follows
\beq\label{glueball-eq-1-3}
  y(1-y)\partial_y^2 P+\left(\gamma-[\alpha+\beta+1]y\right)\partial_x P
          -\alpha\beta P=0\, ,
\eeq
where $y=-x^2$, and
\beq
  \alpha={b\over 2}\, , \quad \beta=2+{b\over 2}\, , \quad \gamma=b-2a-1\, .
\eeq
As for $a$ and $b$, we have four set of solutions,

\vspace{.3cm}
(i) $a={1\over 2}\left( -3-\sqrt{9+\bar{m}^2}\right)$, $b=2a$,

(ii) $a={1\over 2}\left(-3-\sqrt{9+\bar{m}^2}\right )$, $b=2a+4$,

(iii) $a={1\over 2}\left( -3+\sqrt{9+\bar{m}^2}\right )$, $b=2a$,

(iv) $a={1\over 2}\left(-3+\sqrt{9+\bar{m}^2}\right )$, $b=2a+4$,

\vspace{.3cm}
The solution of (\ref{glueball-eq-1-3}) is given by the Gauss's hyper-geometric function
as
\beq
  P(y)=F(\alpha, \beta, \gamma; y)
\eeq
and its behavior is well known. So we obtain the solution in the four forms of
hyper-geometric function. Among them, we find that the solution (ii) and (iv) 
satisfy (\ref{glueball-eq-1-2}). Then the 
other two, (i) and (iii), are not the solution.
Furthermore, we can show that the solution (ii) (denoted by $\phi_2$)  
is equivalent to the one of (iv), then we get only
one solution, $\phi_2$, at this stage.

\vspace{.3cm}
The other independent solution of (\ref{glueball-eq-1-2})
is given as follows.
Set as 
\beq
  \phi_5=Q(x) \phi_2(x)\, ,
\eeq
then from (\ref{glueball-eq-1}), we obtain the equation of $Q(x)$  as
\beq
   \partial_x^2 Q+\left(g_2(x) \phi_2+2\partial_x\phi_2\right)\partial_x Q=0\, .
\eeq
Deviding this Eq. by $\phi_2\partial_x C $, we find
\beq
   \log \left(\phi_2^2\partial_x Q \right)=-\int g_2 dx\, .
\eeq
This is solved as
\beq
    \partial_x Q={\bar{q}_0\over \phi_2^2}{x^3\over (1+x^2)^4}\, .
\eeq
where $\bar{q}_0$ is an integral constant. Finally, we get
\beq\label{sol-Q}
   Q=\int dx {\bar{q}_0\over \phi_2^2}{x^3\over (1+x^2)^4}\, .
\eeq

Here it is possible to add an arbitrary constant to the right hand side of
(\ref{sol-Q}). However it is not necessary since it is absorbed into the coefficient
of $\phi_2$ of general solution. It is given as follows
\beq
   \phi=\alpha_2\phi_2+\alpha_5\phi_5
\eeq
where $\alpha_2$ and $\alpha_5$ are constants. Further, we consider that
$\bar{q}_0$ is absorbed into $\alpha_5$ hereafter.
They are determined by the
boundary conditions on the two boundaries, $r=\infty$ and $r=0$, as follows.

\vspace{.3cm}
\noindent{\bf Glueball (normalizable) solution}

In the limit of $r\to 0$, the above solutions are expanded as
\bea
        \phi_2 &=& x^4\left( 1-{32+\bar{m}^2\over 12} x^2+O\left({x^4}\right)\right)
   \to {0~~~~ }\,  , \\
   \phi_5 &=& -{1\over 4}\left( 1+{\bar{m}^2\over 4} x^2+{(q_4+q_{4L}\ln x) x^4}+\cdots 
                  \right)
   \to {\rm const.~~~~} \, . \label{phi-5-0} \\
   && q_4=-{(8+\bar{m}^2)(32+\bar{m}^2) \over 36}\, ,
\eea 
where $q_{4L}$ is a constant.
 From (\ref{phi-5-0}), we find that $\phi_5$ is non-normalizable since $\phi_5$ is a constant
at $r=0$. This is understood as
\beq
\int dr\sqrt{g_{(5)}}|\phi_5|^2 \sim \int_{r\to 0} dr {1\over r^3}|\phi_5|^2\, ,
\eeq
where the factor $1/r^3$ appears from $\sqrt{g_{(5)}}$ near $r=0$ in the integral measure
of the wave-function. On the other hand, we find
\beq
   \phi_2 \to {\rm const.~~~~}
\eeq
in the limit of $r\to \infty$. This implies that this wave function includes the source
of the field operators of the theory on the boundary $r=\infty$. The normalizable modes
are also found for special values of $m^2$ given below. In this case, it behaves as
\beq
   \phi_2 \to {p_4\over x^4}+\cdots\, 
\eeq
with a constant $p_4$, thus the wave-function $\phi_2$ is normalizable.

Then we take $\phi=\phi_2$ as the wave function for the glueball,
and we find that
this solution is actually normalizable under the condition,
\beq
  \sqrt{9+\bar{m}^2}=5+2n\, , \quad n=0,1,2, \cdots
\eeq 
Thus we get the following mass spectra for the glueball considered here,
\beq\label{mass-formula}
   m^2=4(n+1)(n+4){r_0^2\over R^4}=\lambda (n+1)(n+4)\, , \quad n=0,1,2, \dots\, .
\eeq
This resultant formula is compared with the above formula (\ref{AdS4-mass}).
This coincides with the case of spin two tensor. Namely, the lowest glueball
mass is $m_0=2\sqrt{\lambda}$, which is also obtained here by the WKB
approximation with high accuracy as shown below. 

\vspace{.3cm}
{The analysis for the glueball mass given above is performed for the
theory on the boundary $r=\infty$. In the present case of $b_0=0$, it is parallel to perform
the same analysis for the theory at the boundary $r=0$. The only thing we should do is to change
the variable $r$ to $z=r_0^2/r$, then we will find the same 
mass eigenvalues also in the theory at $z=\infty$. When the dark radiation is added, the symmetric situation is broken and the analysis becomes complicated as shown below.
}

\vspace{.3cm}
\subsection{$0\leq b_0< r_0$ case : WKB approximation}

When finite value of $C$ is introduced, it is impossible to solve 
analytically the equation of motion
for the fluctuation mode of the bulk fields. So we consider here an alternative
method to find the glueball spectra. The most popular and convenient one
is the WKB method which has been used by many people {\cite{Minahan}-\cite{BMT},\cite{GKTT} }.

%%%%%%%%%%%%%%%%%%% FIG %%%%%%%%%%%%%%%%
\begin{figure}[htbp]%[H]
\vspace{.3cm}
\begin{center}
\includegraphics[width=12cm]{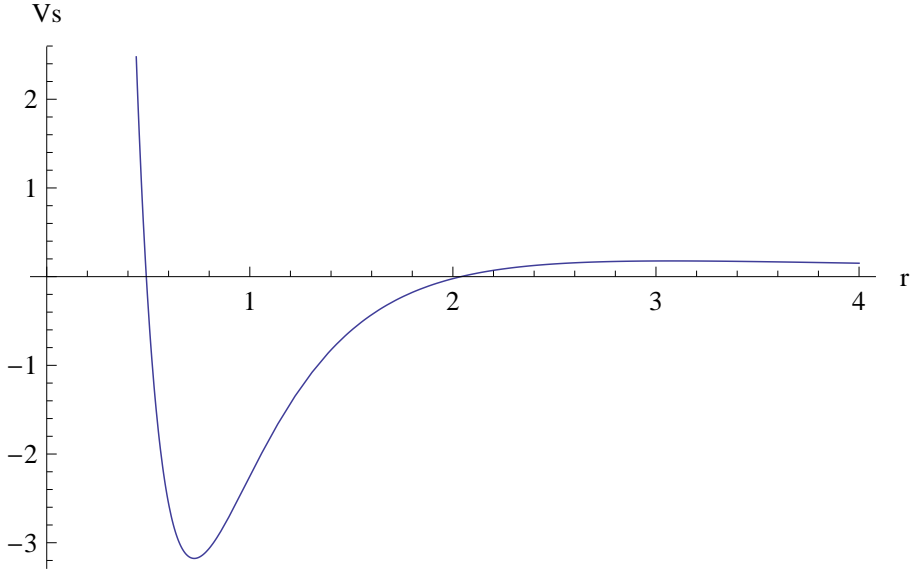}
\caption{{\small The Schr\"{o}dinger potentials $V(r)$ for $r_0=R=1$ and $|\lambda|=4$ 
is shown
for the graviton with $m=2\sqrt{|\lambda|}=4.0$, $r_1=0.4903$ and $r_2=2.0395$. }.
}
\label{NS-Pot-0}
\end{center}
\end{figure}
%%%%%%%%%%%%%%%%%%%%%%%%%%%%%%%%%%%%%%
\vspace{.3cm}
\noindent\underline{\bf $b_0=0$ case}

First, we perform this method to obtain the mass for the case of $C=0$, then its results
are compared with the one given in the previous section to assure that this approximation
is good.

The equation (\ref{glueball-eq-1}) has two regular singularities at
$r=0,~\infty$. Here,
we try to find the eigenfunctions in the region of $0\leq r\leq \infty$ through
WKB approximation \cite{Minahan,COOT}.

\vspace{.3cm}
By factorizing $\phi$ as
\beq\label{wave-eq}
   \phi=e^{-{1\over 2}\int dr g_2(r)}f(r)\, ,
\eeq
the equation (\ref{glueball-eq-1}) is rewritten as
\beq\label{glueball-eq2}
  -\partial_r^2f+V(r)f=0\, , \quad {V={1\over 4}g_2^2+{1\over 2}\partial_rg_2-
{m^2\over A^2 }({R\over r})^4\, }.
\eeq
%\textcolor{red}
This is equivalent to the one dimensional Schr\"{o}dinger equation with the potential $V$
and the zero energy eigenvalue. For an appropriate mass $m$,
we can see that $V$ has two turning points, $r_1$ and $r_2(>r_1)$, to give \cite{Minahan}
\beq\label{WKB-condi}
  \int_{r_1}^{r_2}\sqrt{-V}dr=\left(n+{1\over 2}\right)\pi
\eeq
with an integer $n$. From this equation we obtain the discrete glueball mass $m_n$, where
$n$ denotes the node number of the eigenfunction. The potential for the zero node is
shown in the Fig.\ref{NS-Pot-0}. In this case, we have $\int_{r_1}^{r_2}\sqrt{-V}dr=\pi/2$
and the lowest mass
with 4 percent numerical error compared to the correct eigenvalue obtained through
an analytical solution given above.

\vspace{.5cm}
\noindent\underline{\bf $0<b_0 < r_0$ case}

According to the procedure given above, we find the following equations for this
case for glueball of $2^{++}$.
By setting as $h_{ij}=p_{ij}\chi(x^{\mu})\phi(t,r)$, where 
$\chi(x^{\mu})$ is assume to be satisfied (\ref{4dads}) and $\phi(r)$ is replaced by $\phi(a_0(t),r)$
since the coefficients of the equation are written by using $a_0(t)$.
Then, we get for $\phi(a_0(t),r)$ the following equations,
\bea
   &&\partial_r^2\phi+\bar{g}_2(r)\partial_r\phi+
({R\over r})^4{m^2\over \bar{A}^2 }\phi=J\, ,\label{glueball-eq-12}\\
  && \bar{g}_2(r)=\partial_r\left(\log\left[(r/R)^5\bar{n}\bar{A}^3\right]\right)\, , \label{glueball-eq-22} \\
  && J=-({R\over r})^4{\phi\over \chi\bar{A}^2 }
\left(\partial_t^2+3{\dot{a}_0\over a_0}\partial_t\right)\chi +   
 ({R\over r})^4{1\over \chi\bar{n}^2 }\left(\partial_t^2+3{\dot{a}_0\over a_0}\partial_t\right)
(\chi\phi)-   \nonumber \\
&&  -({R\over r})^4{1\over\chi}\left({\partial_t(\bar{n})\over \bar{n}^3}-3{\partial_t(\bar{A})\over \bar{A}}\right)\partial_t(\chi\phi)\, . \label{glueball-eq-32}
\eea
The left hand side of (\ref{glueball-eq-12}) has a similar form to (\ref{glueball-eq-1-2}), 
so we expect a stable
glueball state. However, the term on the right hand side, $J$, arises 
because of non-zero $C$. In spite of its complicated form, 
$J$ is simplified when $\partial_t(a_0(t))=\dot{a}_0$ is neglected 
according to our approximation adopted above.
In this case, we find
\beq
 J=({R\over r})^4\left({1\over \bar{A}^2 }-{1\over \bar{n}^2 }\right)\Omega^2\phi\, ,
\eeq
where 
\beq
       -\Omega^2 = {\partial_t^2\chi\over \chi}\, .
\eeq

Furthermore, we simplify the situation so that the derivative with respect to the spacial
coordinate for $\chi$ can be neglected. In this approximation, the WKB approximation 
would be usefull especially for the ground state. Then we may set as
\beq
    \Omega^2=m^2
\eeq
and we obtain
\beq\label{finite-C}
     \partial_r^2\phi+\bar{g}_2(r)\partial_r\phi+
({R\over r})^4{m^2\over \bar{n}^2 }\phi=0\, .
\eeq
Notice that the last term of the left hand side is changed from $\bar{A}$ to $\bar{n}$.
This point is the main and important difference from the case of $C=0$. 
Due to this replacement, we find the glueball mass decreases with $C$
and it tends to zero at the transition point $\tilde{c}_0^{1/4}=r_0/R$, where confinement is
lost from.
Of course, this equation
is reduced to (\ref{glueball-eq-1-2}) for $C=0$. 

%\vspace{.3cm}
%\noindent{\bf WKB Approximation}

The equation (\ref{finite-C}) has two regular singularities at
$r=0,~\infty$. Then,
we perform the analysis through WKB approximation in the region of $0\leq r\leq \infty$.
 % \cite{Minahan,COOT}.
%\vspace{.3cm}
By factorizing $\phi$ as
\beq\label{wave-eq}
   \phi=e^{-{1\over 2}\int dr \bar{g}_2(r)}f(r)\, ,
\eeq
the equation (\ref{glueball-eq-1}) is rewritten as
\beq\label{glueball-eq2}
  -\partial_r^2f+\bar{V}(r)f=0\, , \quad {\bar{V}={1\over 4}\bar{g}_2^2
+{1\over 2}\partial_r\bar{g}_2-
{m^2\over \bar{n}^2 }({R\over r})^4\, }.
\eeq
%\textcolor{red}
As shown above for $C=0$ case, for an appropriate mass $m$,
we can see that $\bar{V}$ has two turning points, $r_1$ and $r_2(>r_1)$, which give
%\cite{Minahan}
\beq\label{WKB-condi}
  \int_{r_1}^{r_2}\sqrt{-V}dr=\left(n+{1\over 2}\right)\pi
\eeq
with an integer $n$. 
We show the numerical results for the lowest mass eigenvalue of $n=0$ to see
the effect of the dark radiation $C$.
%In the Fig. we show $\bar{V}$ for several case with $n=0$ and different values of $C$. 

%\newpage
%*********** Nakamura  ********
%\newpage

%*********** Nakamura  ********
\begin{figure}[htbp]%[H]
\vspace{.3cm}
\begin{center}
\includegraphics[width=12cm]{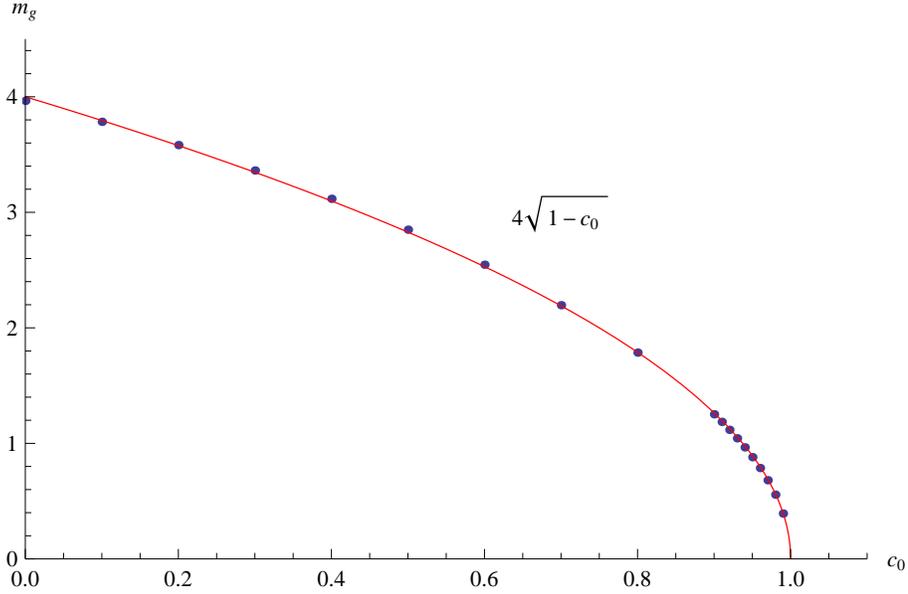}
\caption{{\small  The mass $m_g$ is plotted against $\tilde c_0$.     
The mass $m_g$ is defined by (\ref{glueball-eq2}) and (\ref{WKB-condi}) 
with $n=0$. There $r_0=1$, $R=1$ and $\lambda=-4 r_0^2/R^4 = -4$.  
The fitted curve is given by $m_g=4\sqrt{1- \tilde c_0}$.  
The axis label $c_0$ means $\tilde c_0$.  
}}
\label{tc0m}
\end{center}
\end{figure}

\vspace{.5cm}
\noindent{\bf The results of numerical analysis for $n=0$}

The glueball mass $m_g$ of the ground state is obtained by WKB method 
as mentioned above.  
More specifically, $m_g$ is calculated using (\ref{glueball-eq2}) and 
(\ref{WKB-condi}) with $n=0$.  
Numerical value of $m_g$ is plotted as a function of $\tilde c_0$ in 
Fig. \ref{tc0m}.  The results are well fitted by
\beq
m_g=\sqrt{15.97(1-\tilde c_0)} \approx 4\sqrt{1-\tilde c_0} \, .
\eeq
As expected, $m_g$ decreases and 
{vanishes at the critical point $\tilde c_0=1$.  
%The value of $\tilde c_0$ is 1 at the critical point.   
 }

%%%%%%%%%%%%%%%%%% FIG %%%%%%%%%%%%%%%%

%%%%%%%%%%%%%%%%%%%%%%%%%%%%%%%%%%%%%%
\vspace{.5cm}

{Finally, we give the following comments of WKB analysis given here.

\noindent 1) The dark radiation $C$ is related to 
$\tilde c_0$ as
\beq
\tilde c_0 = C/(4\mu^2 a_0^4)\, .
\eeq
In the present analysis, $a_0$ is assumed to be almost constant.  
And $\mu=1/R$ is fixed to be 1 in the present subsection.  
So the variation of $\tilde c_0$ corresponds to the one of $C$, 
which corresponds to the dark radiation energy density.  
}

\vspace{.3cm}
\noindent 2) It should be worthwhile to mention that there exist solutions of 
(\ref{glueball-eq2}) and (\ref{WKB-condi}) even above the critical point.  
For example, the lightest mass is given by $m_g=0.031$ with $n=3$ 
when $\tilde c_0=1.02$.  However the two turning points of this mode are seen at
$r_1=0.086$ and $r_2=0.252$, and $r_H = 0.100>r_1$ so that $r_1$ is hidden 
behind the horizon.  {In general, all the wave-functions corresponding to
the glueball are not well defined in the region $r_H<r<\infty$ since the region of $r_H>r>0$ is
needed.}
{In other word, these modes might be unstable 
and might be swallowed into the region $r<r_H$ in the final step.  In this sense, they  
correspond to the quasi-normal mode in \cite{ET}.  }

\vspace{.3cm}
\noindent 3) We  should notice that
the glueball mass studied above is the one for the theory at the boundary $r=\infty$.
For the theory at $r=0$, we could see the spectra by using the operator written by $\hat{g}_0$.
This is abbreviated here and in the next section.

\section{Glueballs as Rotating closed string}\label{classical}

In this section, we show the classical
stable configuration of glueballs corresponding to the state of large quantum number.
Then we support the above results for the $C$ dependence of the glueball mass.
The quantum fluctuations around the classical configuration can be neglected in this case.
Flavored mesons are given by an open string whose two end points are on the D7 brane. 
On the other hand, the glueball with higher spin would be represented by a rotating closed string in the bulk.
Such a rotating string is formulated according to \cite{ZSV}-
\cite{Huang:2007fv}, \cite{GTT} as follows.

In performing the analysis, we need only the 5D bulk part of the metric, which is rewritten as
\bea
ds^2_{5}&=&{r^2 \over R^2}\left(-\bar{n}^2dt^2+\bar{A}^2ds^2_{(3)}\right)+
\frac{R^2}{r^2} dr^2 \,  , \label{closed-1} \\
    ds^2_{(3)}  &=&  a_0^2(t)\gamma_{ij}(x)dx^idx^j\, \label{Trans-2} \\
    &=&  a_0^2(t)\left({dp^2\over 1+p^2/\bar{r}^2}+p^2d\Omega^2_{(2)} \right)\, ,
       \label{Trans-3} \\
     d\Omega^2_{(2)} &=& d\bar{\theta}^2+\sin^2\bar{\theta}d\bar{\phi}^2  
\eea
where $\Omega_{(2)}$ denotes the metric of $S^2$ with 
two angle coordinates $\bar{\theta}$ and $\bar{\phi}$, and
\beq
  p={\bar{r}\over 1-{\bar{r}^2\over 4\bar{r_0}^2}}\, \label{Trans-4}
\eeq
Here, we consider a closed string, which rotates around
the podal axis of $S^2$ at a fixed value of $\bar{\theta}$. 

\vspace{.5cm}
\noindent\underline{\bf Ansatz; $p(r)$ and $\phi=\omega t$}

For the simplicity, we consider the solution of the form given by $p(r)$ and $\phi=\omega t$.
In this case, we have the induced metric for the string as
\bea
 {\cal G}_{\tau\tau} &=& {r^2\over R^2}\bar{A}^2(r)\left(
   -{\bar{n}^2\over \bar{A}^2}+\omega^2p^2\sin^2\bar{\theta} a_0^2(t) \right)\, , 
     \label{ind-string-1} \\
 {\cal G}_{\sigma\sigma} &=& {R^2\over r^2}+{r^2\over R^2}\bar{A}^2
                   {{p'}^2\over 1+p^2}a_0^2(t) \, ,
 \label{ind-string-2}
\eea
where $\omega$ is a constant and the prime denotes the derivative with respect to $r$.
Then we have the following Nambu-Goto action for the closed string,
\bea
  S_{\rm string} &=& \int dt {\cal L} \, \label{string-ac0} \\
          &=&
     -{1\over 2\pi\alpha'}\int dtdr %\sqrt{(\dot{X}\cdot{X'})^2-{\dot{X}^2{X'}^2}}
  {r^2\over R^2}\bar{A}^2
        \sqrt{\left({\bar{n}^2\over \bar{A}^2}-\omega^2p^2\sin^2\bar{\theta} a_0^2(t) \right)
      \left({{p'}^2\over 1+p^2}a_0^2(t) +\bar{A}^{-2} \left({R\over r}\right)^4\right)}\, . \nonumber \\
                       \label{string-ac1}
\eea

 From this, the spin $J_s$ and the energy $E_s$ of this string are formally given as
\beq\label{spin}
   J_s={\partial{\cal L}\over \partial\omega}
   ={1\over 2\pi\alpha'}\int dr a_0^2{r^2\over R^2}\bar{A}^2\omega p^2\sin^2\bar{\theta} 
        \sqrt{a_0^2(t){{p'}^2/(1+p^2)} +\bar{A}^{-2} \left({R\over r}\right)^4
           \over {\bar{n}^2/ \bar{A}^2}-\omega^2p^2\sin^2\bar{\theta} a_0^2(t)}\, ,
\eeq
\beq\label{energy}
   E_s=\omega{\partial{\cal L}\over \partial\omega}-{\cal L}
   ={1\over 2\pi\alpha'}\int dr {r^2\over R^2}\bar{n}^2
        \sqrt{a_0^2(t){{p'}^2/(1+p^2)} +\bar{A}^{-2} \left({R\over r}\right)^4
           \over {\bar{n}^2/ \bar{A}^2}-\omega^2p^2\sin^2\bar{\theta} a_0^2(t)}\, .
\eeq
%\subsection

\vspace{.3cm}
\noindent\underline {\bf Solution }
%\noindent{\bf Solution for SUSY background}

\vspace{.3cm}
\noindent\underline{\bf $b_0=0$ case}
\vspace{.3cm}

In this case, we could find a solution of constant $r$ as shown below. 
The Lagrangian is rewritten by supposing $r=r(p)$ as
\beq
  {\cal L} =-
     -{1\over 2\pi\alpha'}\int dp %\sqrt{(\dot{X}\cdot{X'})^2-{\dot{X}^2{X'}^2}}
  {r^2\over R^2}\bar{A}^2
        \sqrt{\left(1-\omega^2p^2\sin^2\bar{\theta} a_0^2(t) \right)
      \left({a_0^2(t)\over 1+p^2} +{\dot{r}^2\over \bar{A}^{2}} \left({R\over r}\right)^4\right)}\, , \nonumber \\
                       \label{s-3}     
\eeq
where $\dot{r}=\partial_p r$. Then imposing $r=$constant ($\dot{r}=0$), the equation of motion
for $r$ is given as
\beq
   \partial_r\left({r^2\over R^2}\bar{A}^2\right)=0\, , \label{eq-r}
\eeq
The solution is found as 
\beq
     r=r_0\, .
\eeq

%\vspace{.5cm}
%\noindent{\bf Regge behavior}

\vspace{.3cm}
The spin and the energy of this closed string configuration are given by using the
above equations (\ref{spin}) and (\ref{energy}) as
\beq\label{spin-2}
   J_s={1\over 2\pi\alpha'} 
      {4a_0l^2\over \omega}{r_0^2\over R^2}\int_{-1/l}^{1/l} dp
        {p^2\over \sqrt{(1+p^2)(1-l^2 p^2)}}\, ,
\eeq
\beq\label{energy-2}
   E_s={1\over 2\pi\alpha'}
      {4a_0}{r_0^2\over R^2}\int_{-1/l}^{1/l} dp 
        {1\over \sqrt{(1+p^2)(1-l^2 p^2)}}\, ,
  \eeq
where $l=\omega a_0\sin \bar{\theta}$.

For small $p$, by approximating as $\sqrt{1+p^2}\sim 1$, we can estimate the above
$J_s$ and $E_s$ as follows.
  
Substituting the above closed string solution, we find
\bea
  J_s &=& {Kl^2\over \omega}\int^{1/l}_{-1/l} dp {p^2\over \sqrt{(1-l^2 p^2)}}
          ={K\pi\over 2l\omega}\, , \\
  E_s &=& {K}\int^{1/l}_{-1/l} dp {1\over \sqrt{(1-l^2 p^2)}}
          ={K\pi\over l}\, . \\
  K &=& {1\over 2\pi\alpha'}{4a_0}{r_0^2\over R^2}\, .
\eea
Then we obtain
\beq
  J_s=\alpha'_{\rm glueball}E_s^2\, , \quad \alpha'_{\rm glueball}
   =\alpha'{R^2\over r_0^2}\sin \bar{\theta}
\eeq
We could find the relation
\beq
   \alpha'_{\rm glueball}={1\over 2}\alpha'_{\rm meson}
\eeq
where $\alpha'_{\rm meson}$ represents the slope parameter of the flavored 
mesons \cite{GTT}.

%\vspace{.5cm}
\vspace{.3cm}
\noindent\underline{\bf $b_0\neq 0$ case}
\vspace{.3cm}

In this case, there is no solution of constant $r$. And it is difficult to find any analytic 
solution of the equations of motion, then we perform numerical analysis in this case.
For the simplicity, we set as $r_0=R=1$ and $a_0=0.5$, 
then the equations are solved by barying the value of $\tilde{c}_0$ in the range of 
$0 \leq \tilde{c}_0 < 1$ to obtain the corresponding solution of $p(r)$. A typical configuration
of the solution is shown in the Fig.~\ref{J-E-C} for $\tilde{c}_0=0.5$ and $\omega=1.0$.

\begin{figure}[htbp]%[H]
\begin{center}
\includegraphics[width=8.0cm,height=4cm]{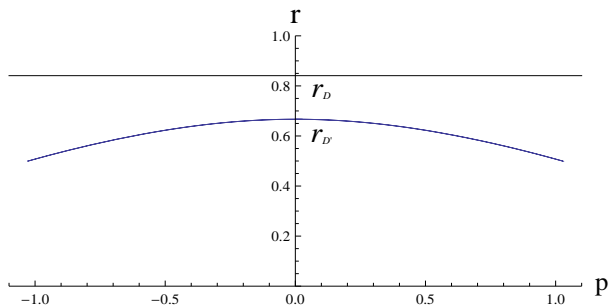}
\caption{The glueball solution $p(r)$ at $\tilde c_0=0.5$ and 
$\omega=1.0$, where $r_D=0.841$.}
\label{J-E-C}
\end{center}
\end{figure}

{We notice that the center of the rotating string ($r_{D'}$) exists in the region of $0<r<r_D$, where 
$r_D$ denotes the domain-wall given above and it separates the bulk to the two
regions corresponding to the two theories which are living in the boundaries at $r=\infty$ and $r=0$. $r_{D'}$ depends on $\omega$. As $\omega \rightarrow 0$,then
 $r_{D'} \rightarrow r_D$. In the present case, the closed string solution appears
in the region corresponding to the boundary $r=0$. So we may consider that the
glueball given here would be observed only in the theory at $r=0$ boundary.
However, as shown in the previous section, the wave function of the glueball extends
in the both region even if the center of the function is at some point in the region
of $0<r<r_D$. In this sense, we could observe the glueball state in the both boundaries.}

\vspace{.3cm}
In the next, we show how this closed string configuration varies with $\tilde c_0$.
Two quantities, (i) the position of its center ($r_{D'}$) and (ii) the length of glueball ($L$),
are examined, and the $\Delta r/r_D =(r_D-r_{D'})/r_D $ and $L$ are shown in the Fig.~\ref{Regge-C}(left).

\begin{figure}[htbp]%[H]
\begin{center}
\includegraphics[width=7.5cm,height=5cm]{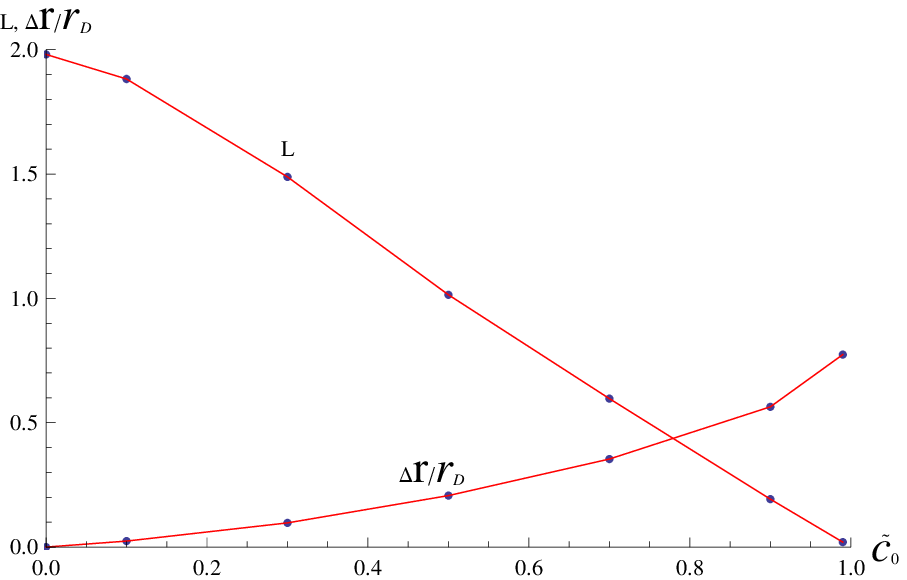}
\includegraphics[width=7.5cm,height=5cm]{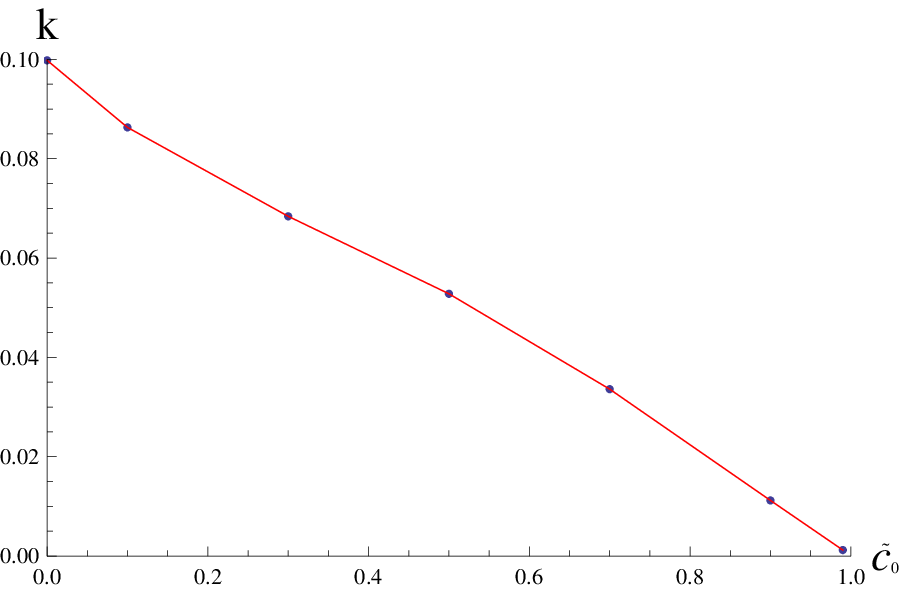}
\caption{Left;The string length $L$ of the glueball and the ratio 
$\Delta r /r_D=(r_D-r_{D'})/r_D$ versus $\tilde c_0$ for $\omega=1.0$.
 $L=\int \sqrt{1+(dr/dp)^2}dp $. 
Right;The string tension $k$ of the glueball versus $\tilde{c}_0$. }
\label{Regge-C}
\end{center}
\end{figure}

 From this figure, we find that the string configuration shrinks to zero size when
$\tilde c_0$ approaches to the critical value, $\tilde c_0 \to 1$. As for the point
$r_{D'}$, it approaches to a point near the boundary $r=0$ but it doesn't touch the boundary.

Further, by changing the value of $\omega$, 
we can see the Regge behaviors 
\beq
J_s=\alpha'_{\rm g}E_s^2\, ,
\eeq 
where $\alpha'_{\rm g}$ depends on $\tilde{c}_0$.  Further, 
$\alpha'_{\rm g}$ is related to the string tension $k$ as 
\beq
\alpha'_{\rm g}=1/(8k)\, .
\eeq  
The results for $k$ are shown in the Fig.~\ref{Regge-C}(right) for various $\tilde{c}_0$.  

The string tension $k$ comes to zero as $\tilde c_0 \to 1$. This is well described
by the line
\beq
   k=0.11(1-\tilde{c}_0)\, .
\eeq
For a state with spin one ($J=1$), we have its mass as
\beq
   m=4\sqrt{1-\tilde{c}_0}\, ,
\eeq
which is the same one obtained in the previous section from WKB approxiamtion.

\vspace{.5cm}
{Through the analyses of this and the previous section, we could see the existence of
glueball state in the confinement region, $0\leq b_0\leq r_0$, and it disappears with
its mass at the critical point $b_0=r_0$. From this, we can say that we can set the order parameter
of this transition as the tension of the QCD string. }

%\newpage
%%%%%%%%%%%%%%%%%%%%%%%%%%%%%%%%%%%%%%%%%%%%%%%%%%%%%%%%%%%%%%%%%%%%%%%%%%%%%%%%%%%%%%%%%%%%%%%%%%%%%%%%%
\section{Entanglement Entropy and Dark Radiation}
%We consider the entanglement entropy for the theory of one boundary. It is given by calculating the minimum area of the surface $A$ whose boundary $\partial A$ is set at the boundary of the bulk.
Here we study the entanglement entropy near the transition region to find an expected sign of
the phase transition.
{As shown in \cite{RT} and \cite{RT2},}  the holographic entanglement entropy is given by 
\begin{equation}\label{see}
S_{EE}=\frac{Area (\gamma_A)}{4G_N^{(5)}} ,
\end{equation}
where $\gamma_A$ denotes the minimal surace whose boundary is defined by $\partial A$ and the surface is extended into the bulk. 
$G_N^{(5)}=G_N^{(10)}/(\pi^3R^5) $ denotes the 5D Newton constant reduced from the 10D one $G_N^{(10)}$.

\vspace{.3cm}
\noindent\underline{ Domain wall for Minimal surface  {$r_c$}:}

 From (\ref{10dmetric-2}),  the spatial part of the bulk metric is rewritten as 
\begin{equation} \label{10dspace-r}
ds_{space}^2=\frac{1}{R^2}\left(r^2+2r_0^2+\frac{r_c^4}{r^2}\right)ds^2_{FRW_3}+\frac{R^2}{r^2}dr^2+R^2d\Omega_5^2 ,
\end{equation}
where
\begin{equation}
ds^2_{FRW_3}=a_0^2(t)\gamma^2(dp^2+p^2d\Omega_2^2) ,
\end{equation}
\begin{equation}\label{gp}
p=\frac{\bar{r}}{\bar{r_0}},\quad\gamma=1/(1-p^2/4) ,
\end{equation}
and $r_c$ is defined as
\begin{equation}
r_c\equiv (\tilde{c}_0R^4+r_0^4)^{1/4} ,
\end{equation}
Notice that, in this section,  $p$ in (\ref{gp}) is different from (\ref{Trans-4}). 
{As mentioned in the section 2, the point $r=r_c$ is called as the domain wall
since the solution of the minimal surface cannot penetrate this point. Namely the solution is restricted to the
region $r_c<r<\infty$ or $0<r<r_c$.}

{As for the part of $0<r<r_c$, the coordinate is rewritten in 
a similar form to the one of $r_c<r<\infty$
by the following change of the variable.}
Change $r$ to $z$ as
\begin{equation} \label{rcz}
z=r_c^2/r ,
\end{equation}
then the spatial part of the bulk metric (\ref{10dspace-r})  becomes   
\begin{equation} \label{10dspace-z}
ds_{space}^2=\frac{1}{R^2}\left(z^2+2r_0^2+\frac{r_c^4}{z^2}\right)ds^2_{FRW_4}+\frac{R^2}{z^2}dz^2+R^2d\Omega_5^2 .
\end{equation}
{It is obvious that the solution would be obtained in the side
$z_c<z<\infty$ in the same form with the one given for $r_c<r<\infty$ by changing
$r$ to $z$.}
Thus it is convenient to use the transformation (\ref{rcz}). 
In $\tilde{c}_0<r_0^2/R^2$,   there is a domain wall at $z=r_c$.

%This is the same form as (\ref{Trans-2}) when $z$ is changed to $r$ where we set $\bar{r}=1$.

\vspace{.3cm}
\noindent{\bf Minimal surface configuration}

Here we consider $\gamma_A$ as a ball with the radius $p_0$ which is fixed at $z=0$.
Then the area of the minimal surface with this boundary $\gamma_A$ is given by
%\begin{equation} \label{sarea}
%\frac{S_{Area}}{4\pi}=\int d\Omega_2\int_0^p dp p^2B\sqrt{B+\frac{R^2z'(p)^2}{z(p)^2}}
%\end{equation}
\begin{equation} \label{sarea}
\frac{S_{Area}}{4\pi}=\int_0^{z(p=0)} dz \mathcal{L}(z) , 
\end{equation}
where
\begin{equation} \label{lz}
\mathcal{L}(z)\equiv p(z)^2B\sqrt{Bp'(z)^2+\frac{R^2}{z^2}} ,
\end{equation}
and 
\begin{equation}
B\equiv \frac{a_0^2\gamma^2}{R^2}\left(z^2+\frac{r_c^4}{z^2}+2r_0^2\right) ,
\end{equation}
%%%%%%%%%%%%%%%%%%%%%%%%%%%%%%%%%%%5
\begin{figure}[htbp]
\vspace{.3cm}
\begin{center}
\includegraphics[width=7.0cm,height=7cm]{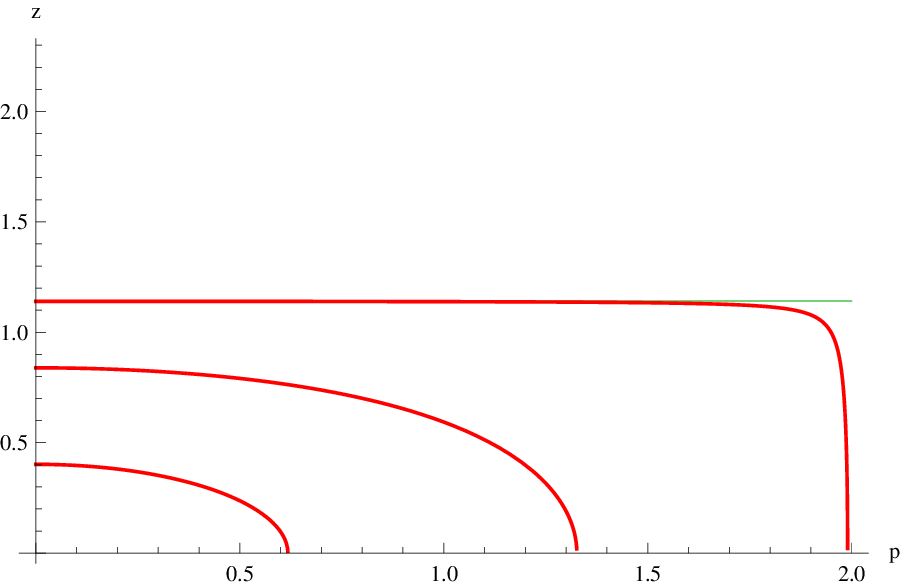}
\includegraphics[width=7.0cm,height=7cm]{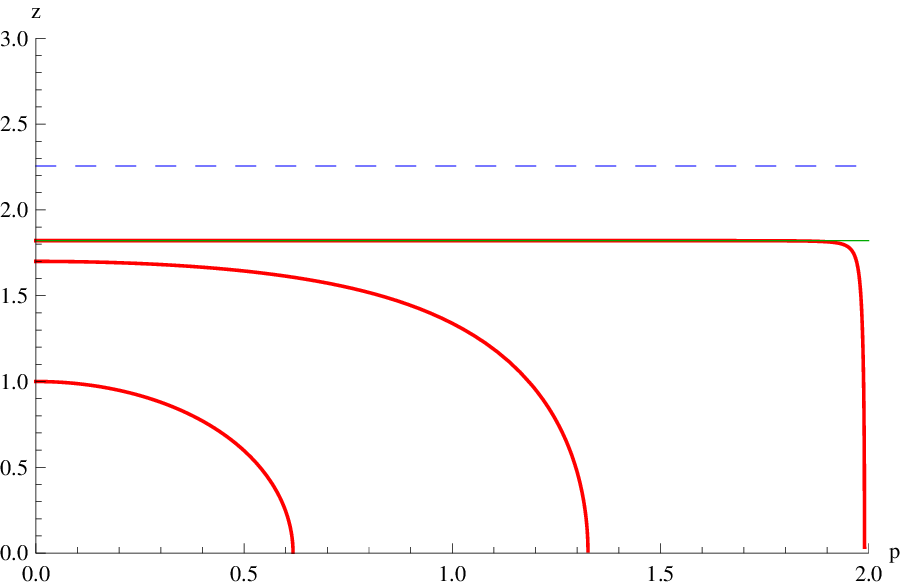}
\caption{{\bf Left;} Embedded solutions for $p(z)$ for $p_0=0.62$, $1.33$ and $1.99$ with $r_0=R=1, a_0=0.5$, $\tilde{c}_0=0.7$. The green line is the  domain wall  $r_c=1.14$. {\bf Right;}
 Embedded solutions for $p(z)$ for $p_0=0.62$ and $1.32585 $ with $r_0=R=1, a_0=0.5$, $\tilde{c}_0=10$. The green line is the domain wall $r_c=1.82$ and the dashed blue line is the event horizon $z_H=2.26$\label{pz10}}
\end{center}
\end{figure}
%%%%%%%%%%%%%%%%%%%%%%%%%%%%%%%%%5

By solving the variational equations from (\ref{sarea}), we can get the minimal surface $p(z)$. The numerical solutions for confinement phase ($c_0<R^4/r_0^4$) and deconfinement  phase ($R^4/r_0^4\le \tilde{c}_0$)  are shown  in Fig.\ref{pz10} where $p_0$ denotes the ball radius on $\gamma_A$
\begin{equation}\label{bound}
 p_0\equiv p(z=0) \leq 2.
\end{equation} 
The upper bound comes from its definition.

\vspace{.2cm}
In the confinement phase, the solutions for $p(r)$ at small $r$ are obtained in the same form
with the one given in the left hand side of the Fig.\ref{pz10} by replacing $z$ by $r$. 

On the other hand,
in deconfinement phase, horizon ($z=z_h\equiv r_c^2/r_h$) appears in the small $r$ side from
the domain wall ($z=r_c$) as shown in the right hand side of the Fig.\ref{pz10}. This relation
is understood from
\begin{equation}
z_H^4-r_c^4=2\sqrt{\bar{c}_0}R^2r_0^2\left(\frac{z_h}{r_c}\right)^4\ge 0 ,
\end{equation}
then the domain wall $r_c$ is  smaller than the horizon $z_h$. The solutions of small $z$
side could not pass the domain wall, then they are away from the horizon. However, the solutions
in the small $r$ side, which are obtained by replacing $z$ by $r$ in the right of the Fig.\ref{pz10}, then
the horizon is given by $r_H=r_c^2/z_H=1.46$. Then the upper two solutions 
for $p_0=1.33$ and $p_0=1.99$ shown in the figure
cross the horizon. When we reject such solutions as acausal one, the upper bound (\ref{bound})
is modified by the value depending on $\tilde{c}_0$. In this sense, the phase transition is
reflected in the theory at boundary $r=0$. 

\vspace{.3cm}
In the next, we try to find the sign of the phase transition in the theory at $z=0$.
In this case, we use the bound (\ref{bound}) at any value of $\tilde{c}_0$.

\vspace{.3cm}
\noindent{\bf Asmptotic solution for $p(z)$ and divergent terms}

The solution $p(z)$ is expanded around $z=0$ as
\begin{equation}\label{pzuv}
p=p_0+p_2z^2+p_4z^4+p_{4L}z^4\log z\cdots ,
\end{equation}
where $p_0=p(z=0)$ and $p_4$ are arbitary constants. $p_2$ is determined as
\begin{equation}\label{p2}
p_2=-\frac{(1-(p_0^2/4)^2)R^4}{2a_0^2p_0r_c^4}\, ,
   \quad p_{4L}=-\frac{\left(1-(p_0^2/4)^2\right)R^8\dot{a}_0^2}{4a_0^4p_0r_c^8} ,
\end{equation}
%\textcolor{red}{and
%\begin{equation}\label{p4L}
%p_{4L}=-\frac{\left(1-(p_0^2/4)^2\right)R^8\dot{a}_0^2}{4a_0^4p_0r_c^8} ,
%\end{equation}}
where we used {(\ref{bc-3-1})} with $k=-1$. 
{When the time dependence of $\dot{a}_0(t)/a_0(t)$ is small ($\dot{a}_0\sim 0$),  $p_{4L}\sim 0$} .

By using  (\ref{pzuv}) and (\ref{p2}), the integrand (\ref{lz}) is expanded around $z=0$ as
\begin{equation}
\mathcal{L}(z)=\frac{16a_0^2p_0^2r_c^4}{(p_0^2-4)^2R}\frac{1}{z^3}+\frac{64a_0^2p_0^2r_0^2-(p_0^2+4)^2}{2(p_0^2-4)^2R}\frac{1}{z}+\mathcal{O} (z) .
\end{equation} 
Then, {area of the minimal surface} of the region with radius $p=p_0$ is given by 
\begin{eqnarray}
&\frac{S_{Area}}{4\pi}&=\int_\epsilon^{z(p=0)} dz \mathcal{L}(z)\\
&=&\frac{8a_0^2r_c^4p_0^2}{(p_0^2-4)^2R}\frac{1}{\epsilon^2}-\frac{64a_0^2p_0^2r_0^2-(
p_0^2+4)^2R^4}{2(p_0^2-4)^2R}\log\left(\frac{\epsilon}{p_0}\right) + S_{finite} ,
\end{eqnarray}
 where the first and second term is the  UV ($\epsilon \to 0$) divergent terms, and $S_{finite}$ is a finite terms for UV limit ($z=\epsilon\to 0$) .
{Then, from  (\ref{see}) and the relation $R^4=4\pi g_s\alpha'^2N$ ,  
the entanglement entropy becomes 
\begin{equation}
S_{EE}=\gamma_1\frac{Area(\partial A)}{4\pi\epsilon^2}+\gamma_2\log\left(\frac{p_0}{\epsilon}\right)\cdots
\end{equation}
\begin{eqnarray}
\gamma_1&=&\frac{N^2r_c^4}{R^4}\,  , \\ 
\gamma_2&=&N^2\left(1+\frac{Area(\partial A)}{4\pi}\left(\frac{\dot{a}_0}{a_0}\right)^2\right) \label{gm2} ,
\end{eqnarray}
where $k_0=p_0/(1-p_0^2/4)$
and $ Area(\partial A)$ denotes  the proper area of the surface $A$ which is defined as 
\begin{equation}
Area(\partial A) =  4\pi k_0^2 a_0(t)^2
\end{equation}
{The second term of (\ref{gm2}) is the effect of the curvature of $FRW_4$\cite{Mal-ds}. }

When the time dependence of $a_0(t)$ is small ($\dot{a}_0\sim 0$),  $\gamma_2\sim N^2$ which is  the degree of freedoms  in the dual field theory .} 

\vspace{.3cm}
\noindent{\bf Behavior of the finite part $S_{finite}$}
 
In the next, we observe the behavior of the finite part $S_{finite}$ of the entanglement entropy.
On the boundary $z=0$, this quantity is calculated by using the common formula in all the range
of $\tilde{c}_0$.

\begin{figure}[htbp]%[H]
\begin{center}
\includegraphics[width=14.0cm,height=7cm]{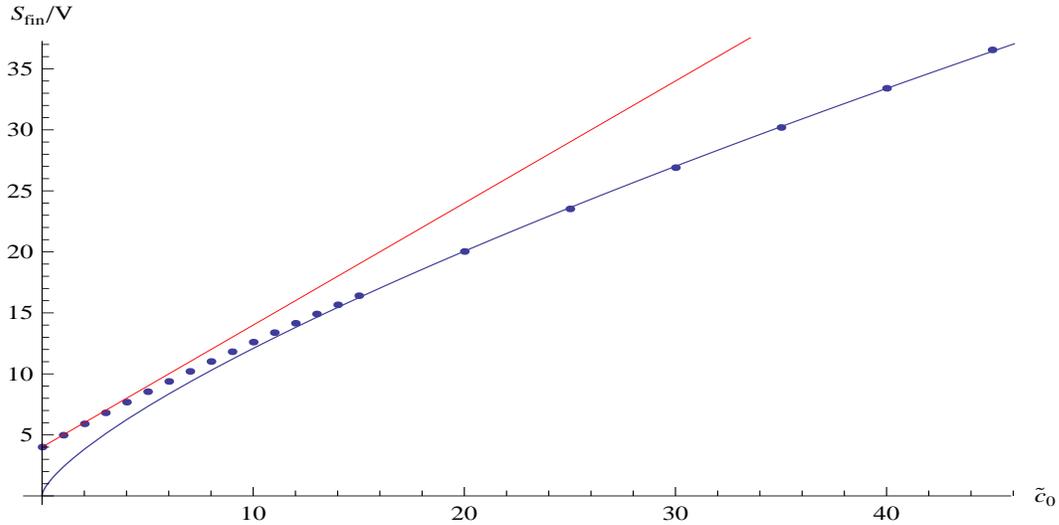}
\caption{A dotted line is $S_{finite}/V$ with $\tilde{c}_0$ at $p_0=1.99, V=2488, a_0=0.5, R=r_0=1$. $S_{finite}$ can be fitted by $S_{finite}/V=0.97\tilde{c}_0+4.01$ at small $c_0$ region and $S_{finite}/V=2.05\tilde{c}_0^{3/4}+0.32\tilde{c}_0^{1/4}$ at large $c_0$ region respectively}
\label{sfin-total}
\end{center}
\end{figure}

In the Fig. \ref{sfin-total}, $S_{finite}/V$ is shown for $p_0=1.99, V=2488, a_0=0.5, R=r_0=1$. 
Here, $S_{finite}$ is normalized by 
the volume  $V$ of the sphere  with radius $p=p_0$ in $FRW_4$ space (\ref{10dspace-z}).
It is given as 
\begin{equation}
V=a_0^3\int_0^{p_0}\gamma^3p^2dp=\frac{1}{2}a_0^3\left(\frac{4p_0(4+p_0^2)}{(p_0^2-4)^2}+\log\frac{2-p_0}{2+p_0}\right) .
\end{equation}  
 From the figure, we can't see any abrupt change near the transition point. 
However, its 
$\tilde{c}_0$ dependence changes from the small to the large $\tilde{c}_0$. 
For small $\tilde{c}_0$ region,
\beq
   S_{finite}/V=0.97\tilde{c}_0+4.01
\eeq
 and  large $\tilde{c}_0$ region.
\beq\label{temp-dep}
   S_{finite}/V=2.05\tilde{c}_0^{3/4}+0.32\tilde{c}_0^{1/4}.
\eeq 
This implies that the increasing behaviour of the entropy in the two regions 
seems to be dominated by different dynamical origin in each region. The transition
from the one at small $\tilde{c}_0$ to the larger one seems smooth.
The values are obtained at $p_0=1.99$. This means the
entanglement entropy is estimated for large volume limit. In this case, we would expect
that the entanglement entropy approaches to the usual thermal entropy of the system
when it has temperature.  In the present case, the first term of (\ref{temp-dep}) indicates
$S\propto T^3$, the behavior of the thermal entropy with the temperature $T$.

Actually, in deconfiniment phase($\tilde{c}_0\ge \frac{r_0^4}{R^4}$),  
the Hawking temperature $T_h$ 
appears with the event horizon at $r_H$, and it is calculated as Appendix. (\ref{Th}). Then we plot $T_h$ dependence of finite part of the entanglement entropy $S_{finite}$ for $\tilde{c}_0>1$ in Fig.\ref{fig-ths}. As expected, we find the behavior, $S_{finite} \propto T_h^3$ 
for the region of large temperature {as shown in \cite{Swingle:2011}}. This point is assured by 
the behavior of the thermal entropy which is shown in Appendix B. 
%For region of  small temperature, $S_{finite}\propto T_h^2$.

In the confiniment phase ($\tilde{c}_0< \frac{r_0^4}{R^4}$), however, $S_{finite}$ 
increases with $\tilde{c}_0$ linearly as shown in Fig.\ref{sfin-total}. This behavior will be
discussed in the future.

%%%%%%%%%%%%%%%%%%%%%%%%%%%%%%%%%%%5
\begin{figure}[htbp]%[H]
\vspace{.3cm}
\begin{center}
\includegraphics[width=14.0cm,height=7cm]{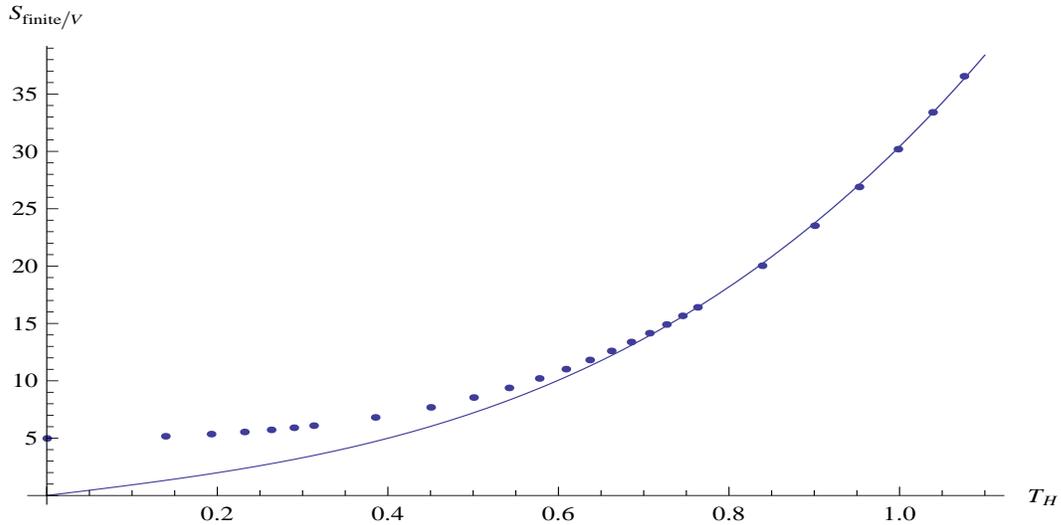}
\caption{A dotted line is $S_{finite}/V$ with $T_h$ at $p_0=1.99, a_0=0.5, V=2488,  R=r_0=1$. $S_{finite}$ can be fitted by $S_{finite}=20.4T_h^3+9.06T_h$ for large $T_h$.  
\label{fig-ths}}
\end{center}
\end{figure}
%%%%%%%%%%%%%%%%%%%%%%%%%%%%%%%%%

%\newpage

\section{Summary and Discussion}

We have examined the gravity dual of
the SYM theory in the FRW type space-time, which
is controlled by two essential ingredients,
the 4D cosmological constant $\lambda$ and the dark radiation $C$.
For negative $\lambda$ and $C=0$, the SYM theory is in the confinement phase. 
On the other hand, the theory is in the deconfinement phase with finite temperature
for $\lambda=0$ and finite $C$. This implies that
the dark radiation works as a thermal bath of the SYM system.
When both the negative $\lambda$ and $C$ are existing at the same time, 
they compete each other, and we can observe the phase transition from the confinement to the
deconfinement phase when the value of $C$ increases from very small value.

Here, through the glueball spectra and the entanglement entropy, we have studied 
how this phase tansition is observed by varying the magnitude of $C$. 
As for the glueball, we could show the exact form
of the glueball spectra in the case of $C=0$ by solving the equation
analytically, and we could assure that
the result is consistent with the one of the free fields in the AdS$_4$ space-time.
The latter has been given  in the field theory  
many years ago \cite{AIS,Fr}. 

When the dark radiation $C$ is added in this system, it becomes difficult
to obtain the analytical result. Then we adopted WKB approximation and examined the
the lowest glueball mass numerically, and we find that the mass decreases with increasing 
dark radiation in the region of confinement phase.
Then, near the critical point, the mass of the glueball seems to be vanishing. This behavior
is also observed by solving the classical closed string state in the bulk. This is corresponding
to the high mass glueball state with higher spin. In this analysis,
however, we must be carefull in the region of small mass which is realized near the critical
point since the quantum corrections would be important in this region. In any case, the glueball
state disappears above the critical value of the dark radiation. Then the system moves to
the high temperature deconfinement phase, where the temperature is given by the Hawking
temperature whose value is determined by the dark radiation $C$.

In the analysis of glueball, we give a comment related to the two boundaries which exist
in the confinement region. The two theories on each boundary are symmetric at $C=0$, 
and then the mass spectra are the same with each others. However, when the dark radiation
is added, it works differently in the two boundaries. As a result, we find two different  
theories for $C\neq 0$. Here, we have examined the mass $m$, which is defined in the 
theory at the boundary $r=\infty$. It would be possible to see the mass defined on the 
boundary $r=0$ by using $\hat{g}^{00}m^2$ and $z=r_0^2/r$. To study more on this point
is postponed as a future work.

As for the entanglement entropy, its behavior is described in a symmetric form in both boundaries
in the confinement phase. However, in the deconfinement phase or large $C$ region, 
in the theory on the boundary $r=0$, the 
size of the connected minimal surface is restricted by the value of $C$ when the surface
is restricted to the causal region. In other words, the large sized surface is disconnected
since the small sized part is cut off by the horizon.

So we have examined the entanglement entropy observed from $r=\infty$ boundary in order to
find a sign of the phase transition near the critical point. While we could not find
a clear transition sign at the critical point, we could observe thermal entropy for large
volume area in the deconfinement phase. The entanglement entropy grows like $T_H^3$ 
at large $T_H$. On the other hand, in the confinement phase,
the entanglement entropy increases with the dark radiation linearly. On this point, we will
discuss in the future.

\vspace{.3cm}
%%%%%%%%%%%%%%%%%%%%%%%%%%%%%%%%%%%%
\section*{Acknowledgments}
{The work of M. Ishihara was supported by World Premier International Research Center Initiative WPI, MEXT, Japan. M. I. thanks to participants of  YIPQS workshops "Holographic vistas on Gravity and Strings"  for useful discussions.} 

\newpage

%\newpage

%%%%%%%%%%%%%%%%%%%%%%%%%%

%%%%%%%%%%%%%%%%%%%%%%%%%%%
\def\theequation{A. \arabic{equation}}
\setcounter{equation}{0}

\appendix

\noindent{\bf\Large Appendix}

\section{Brief review of the model}

First, we briefly review our model \cite{EGR,EGR2,GN13}.
We start from the 
10d type IIB supergravity retaining the dilaton
$\Phi$, axion $\chi$ and selfdual five form field strength $F_{(5)}$,
\beq\label{2Baction}
 S={1\over 2\kappa^2}\int d^{10}x\sqrt{-g}\left(R-
{1\over 2}(\partial \Phi)^2+{1\over 2}e^{2\Phi}(\partial \chi)^2
-{1\over 4\cdot 5!}F_{(5)}^2
\right), \label{10d-action}
\eeq
where other fields are neglected since {we do not need} them, and 
$\chi$ is Wick rotated \cite{GGP}.
Under the Freund-Rubin
ansatz for $F_{(5)}$, %the five form field strength, 
$F_{\mu_1\cdots\mu_5}=-\sqrt{\Lambda}/2~\epsilon_{\mu_1\cdots\mu_5}$ 
\cite{KS2,LT}, and for the 10d metric as $M_5\times S^5$,
$$ds^2_{10}=g_{MN}dx^Mdx^N+g_{ij}dx^idx^j=g_{MN}dx^Mdx^N+R^2d\Omega_{5}^2\, ,$$ 
we consider the solution. Here, the parameter is set as $(\mu=)1/R=\sqrt{\Lambda}/2$.

While the dilaton $\Phi$ and the axion $\chi$ play an important role when the bounadary
of $M_5$ is given by Minkowski space-time \cite{KS2,LT,background}, 
we neglect them here since we study the case
of (A)dS$_4$ boundary.
Then the equations of motion of non-compact five dimensional part
$M_5$ are written as
\footnote{The five dimensional $M_5$ part of the
solution is obtained by solving the following reduced 
Einstein frame 5d action,
\beq\label{action}
 S={1\over 2\kappa_5^2}\int d^5x\sqrt{-g}\left(R+3\Lambda
\right), \label{5d-action}
\eeq
which is written 
in the string frame and taking $\alpha'=g_s=1$ and the opposite sign
of the kinetic term of $\chi$ is due to the fact that
the Euclidean version is considered here \cite{GGP}.}

\beq\label{gravity}
 R_{MN}=-\Lambda g_{MN}\, .
\eeq
While this equation leads to the solution of Ad$S_5$, there are various Ad$S_5$ forms of the
solutions which are discriminated by the geometry of their 4D boundary as shown below.

%%%%%%%
\subsection{Solution}\label{sec22}

A class of solutions of the above equation (\ref{gravity})
are obtained in the following form of metric \cite{GN13}, 
\beq\label{10dmetric-2}
ds^2_{10}={r^2 \over R^2}\left(-\bar{n}^2dt^2+\bar{A}^2a_0^2(t)\gamma_{ij}(x)dx^idx^j\right)+
\frac{R^2}{r^2} dr^2 +R^2d\Omega_5^2 \ . 
%\label{finite-c-sol-3}
\eeq
where
\beq\label{AdS4-30} 
    \gamma_{ij}(x)=\delta_{ij}\left( 1+k{\bar{r}^2\over 4\bar{r_0}^2} \right)^{-2}\, , \quad 
    \bar{r}^2=\sum_{i=1}^3 (x^i)^2\, ,
\eeq
and $k=\pm 1,$ or $0$. The arbitrary scale parameter  $\bar{r_0}$ is set hereafter as $\bar{r_0}=1$.
For the undetermined non-compact five dimensional part, the following equation is obtained from the
$tt$ and $rr$ components of (\ref{gravity}) \cite{BDEL,Lang},
\beq\label{A1}
 \left({\dot{a}_0\over a_0}\right)^2+{k\over a_0^2}=
   -{\Lambda\over 4}A^2+\left({{r\over R}A'}\right)^2
  +{C\over a_0^4 A^2}\ , 
\eeq
where $\dot{a_0}=\partial a_0/\partial t$, $A'=\partial A/\partial r$, and 
\beq
 A={r\over R}\bar{A} , \quad {\partial_t({a_0(t)A})\over \dot{a}_0(t)}={r\over R}\bar{n}\, .
\eeq
The constant $C$ is given as an integral constant in obtaining (\ref{A1}), 
and we could understand that it corresponds to
the thermal excitation of ${\cal N}=4$ SYM theory for $a_0(t)=1$, and  
{it is called as dark radiation \cite{BDEL,Lang}.  

At this stage, two undetermined functions, $\bar{A}(r,t)$ and $a_0(t)$, are remained.
{However} the equation to solve
them is the Eq.(\ref{A1}) only.  {Therefore, we could determine $a_0(t)$ 
by introducing the 4D Friedmann equation, 
which is independent of (\ref{gravity}). However it should be realized on the boundary where
various kinds of matter could be added in order to form the presumed FRW universe as in \cite{GN13}}
\bea\label{bc-RW2}
  \left({\dot{a}_0\over a_0}\right)^2+{k\over a_0^2}&=& %\mathop{=}_{y\to\infty} 
 {\Lambda_4\over 3}+{\kappa_4^2\over 3}\left(
    {\rho_m\over a_0^3}+ {\rho_r\over a_0^4}+{\rho_u\over a_0^{3(1+u)}} \right)\equiv
        \lambda(t)\, \label{bc-RW3}
\eea
where $\kappa_4$ ($\Lambda_4$) denotes the 4D gravitational constant 
(cosmological constant).  The quantities $\rho_m$ 
and $\rho_r$ denote the energy density of 
the nonrelativistic matter and the radiation of 4D theory respectively. 
The {most right hand side} expression $\lambda(t)$ 
in (\ref{bc-RW3}) is given as a simple form of
the {most} left hand side of (\ref{bc-RW3}) given by 
using $a_0(t)$. Then the remaining
function $A(t,r)$ is obtained from (\ref{A1})
in terms of $\lambda(t)$. {The last term $\rho_u$ { in the middle of
(\ref{bc-RW3})} represents an unknown matter
with the equation of state, $p_u=u \rho_u$, where $p_u$ and 
$\rho_u$ denote {its} pressure and energy 
density respectively.}
%represents the same one of (\ref{lambda}) given above although some terms are written explicitly in the 
%equation (\ref{bc-RW3}). 
It is important to be able to solve the bulk equation (\ref{A1}) in this way by relating
its left hand side to the Friedmann equation defined on the boundary \cite{GN13} since
we could have a clear image for the solution.

Finally, %in this section, we give
the solution %of $A$ and $n$ by using $\lambda(t)$. They 
is obtained as
%by replacing the coordinate from $y$ to $r$ defined as$r/R=e^{\mu y}$. Then, from (\ref{eqn})-(\ref{5d-metric-2}), we have 

\bea
 \bar{A}&=&\left(\left(1-{\lambda\over 4\mu^2}\left({R\over r}\right)^2\right)^2+\tilde{c}_0 \left({R\over r}\right)^{4}\right)^{1/2}\, , \label{sol-10} \\
\bar{n}&=&{\left(1-{\lambda\over 4\mu^2}\left({R\over r}\right)^2\right)
         \left(1-{\lambda+{a_0\over \dot{a}_0}\dot{\lambda} \over 4\mu^2}\left({R\over r}\right)^2\right)-\tilde{c}_0 \left({R\over r}\right)^{4}\over 
       \sqrt{\left(1-{\lambda\over 4\mu^2}\left({R\over r}\right)^2\right)^2+\tilde{c}_0 \left({R\over r}\right)^{4}}}\, , \label{sol-11}
\eea
where 
\beq
\tilde{c}_0=C/(4\mu^2a_0^4)\, . \label{sol-12}
\eeq

\section{Thermal entropy}
\def\theequation{B. \arabic{equation}}
\setcounter{equation}{0}

In the deconfinement phase, horizon appears at $r=r_H=\sqrt{b_0^2-r_0^2}$, then the
Hawking temperature $T_H(b_0)$ in this case is obtained as
\beq \label{Th}
   T_H(b_0)={r_H\left(1+{r_0^2+b_0^2 \over r_H^2}\right)\over \pi R^2\bar{A}(r_H) }\, ,
\eeq
which approaches to $T_H$ given by (\ref{temperature-0}) for $r_0\to 0$. Then the 
Euclidean action in this case is estimated as
\bea
  \beta F=I &=& {1\over 2\kappa_5^2}\int d^5x\sqrt{-g}\left(R+3\Lambda
\right)\,  \nonumber \\
 &=& -{\Lambda\over 2\kappa_5^2} {V_3\over T_H(b_0)}\int_{r_H}^{\infty} dr
                 \left({r\over R}\right)^3\bar{n}\bar{A}^3\,  ,  \label{5d-action} \\
  V_3 &=& 4\pi a_0^3\int dp {p^2\over \sqrt{1+p^2}}\, \label{3volume}
\eea
where $V_3$ denotes the three dimensional volume of the thermal system.

Then the regularized free energy $F$ is obtained as follows
\beq
  \int_{r_H}^{\infty} dr
                 \left({r\over R}\right)^3\bar{n}\bar{A}^3=ar_H^4+br_H^2+O(r_H^0)\, ,
\eeq
where the coefficients $a,~b$ are written by $r_0$ and $R$. Then we can see at
large $b_0$
\beq
    F\propto T_H^4\, , \quad {\rm then} \quad S\propto T_H^3
\eeq
as in the case of $r_0=0$, namely in the Minkowski space-time case.

%%%%%%%%%%%%%%  References %%%%%%%%%%%%%%

\newpage
\end{document}